\documentclass[a4paper,11pt]{article} 

\usepackage[T1]{fontenc}               
\usepackage{lmodern}                   
\usepackage{microtype}                 
\usepackage{amsmath, amssymb, amsfonts} 
\usepackage{graphicx}                  
\usepackage{hyperref}                  
\usepackage{geometry}                  
\usepackage{cite}                      
\usepackage{authblk}                   
\usepackage{booktabs}                  
\usepackage{caption}
\usepackage{subcaption}                
\usepackage{enumitem}                  

\usepackage[x11names]{xcolor}
\usepackage{float}
\usepackage{soul} 
\usepackage{cleveref}
\crefname{figure}{fig.}{figs.}
\Crefname{figure}{Fig.}{Figs.}
\hypersetup{
    colorlinks=true,
    linkcolor=blue,
    citecolor=blue,
    urlcolor=blue
}
\def\beq{\begin{equation}}
\def\eeq{\end{equation}}
\def\bea{\begin{eqnarray}}
\def\eea{\end{eqnarray}}
\def\beqn{\begin{eqnarray}} 
\def\eeqn{\end{eqnarray}}
\def\nn{\nonumber}
\def\Eq#1{Eq.~(\ref{#1})}

\def\qon#1{q_{#1,0}^{(+)}}

\def\ket#1{|{#1}\rangle}

\def\qb{\mathbf{q}}

\def\ii{\imath 0}

\pdfoutput=1

\newcommand{\toff}[0]{\texttt{t}}
\newcommand{\Toff}[0]{\texttt{T}}\newcommand{\ph}[1]{\phantom{#1}}

\newcommand{\valencia}{Instituto de F\'{\i}sica Corpuscular, Universitat de Val\`{e}ncia 
-- Consejo Superior de Investigaciones Cient\'{\i}ficas, 
Parc Cient\'{\i}fic, E-46980 Paterna, Valencia, Spain.}

\newcommand{\culiacanA}{Facultad de Ciencias F\'{\i}sico-Matem\'aticas,
Universidad Aut\'onoma de Sinaloa, Ciudad Universitaria, CP 80000 Culiac\'an, Mexico.}

\title{\bfseries Quantum querying based on multicontrolled Toffoli gates for causal Feynman loop configurations and directed acyclic graphs}
\author[1,2]{Selomit Ramírez-Uribe}
\author[1]{Andŕes E. Rentería-Olivo}
\author[1]{Germán Rodrigo}
\affil[1]{\valencia}
\affil[2]{\culiacanA}
\date{\today}

\begin{document}

\setcounter{page}{1}
\maketitle

\begin{abstract}
    Quantum algorithms are a promising framework for unfolding the causal configurations of multiloop Feynman diagrams, which is equivalent to querying the \textit{directed acyclic graph} (DAG) configurations of undirected graphs in graph theory. In this paper, we present a quantum algorithm for querying in both types of applications, using a systematic and sparing logic in the design of an oracle operator. The construction of the quantum oracle is based exclusively on multicontrolled Toffoli (MCX) gates and quantum NOT (Pauli-$X$) gates. The efficiency of the algorithm is evaluated by comparison with a quantum algorithm based on binary clauses. Furthermore, we analyse the impact of traspilation and introduce an appropriate metric to assess the complexity of the algorithm, the \emph{quantum circuit area}. We explicitly analyse three-, four- and five-eloop topologies, which have not previously been explored due to their higher complexity and the current limitations of quantum simulators.
\end{abstract}

\section{Introduction}
\label{sec:Introduction}

Quantum computing~\cite{Feynman:1981tf} is currently an appealing approach with great potential to tackle complex problems in many fields where exploiting the quantum principles of superposition and entanglement could suppose an advantage. In the context of high-energy physics~\cite{Humble:2022klb,Rodrigo:2024say} recent applications of quantum algorithms include lattice gauge theories~\cite{Jordan:2011ne,Banuls:2019bmf,Zohar:2015hwa,Byrnes:2005qx}, track reconstruction~\cite{Magano:2021jzd,Duckett:2022ccc,Schwagerl:2023elf}, jet clustering algorithms~\cite{Wei:2019rqy,Pires:2020urc,deLejarza:2022bwc,deLejarza:2022vhe}, analysing the formation of jets in a medium~\cite{Barata:2021yri,Barata:2022wim,Barata:2023clv}, simulation of parton showers~\cite{Bauer:2019qxa,Bauer:2021gup,Bepari:2020xqi,Williams:2021lvr}, determination of parton densities~\cite{Perez-Salinas:2020nem,Cruz-Martinez:2023vgs}, helicity amplitudes~\cite{Bepari:2020xqi} and the colour algebra~\cite{Chawdhry:2023jks} of elementary processes, heavy-ion collisions~\cite{deJong:2020tvx}, quantum machine learning~\cite{Guan:2020bdl, Wu:2020cye,Trenti:2020ceh}, and quantum integrators~\cite{Herbert:2021xgs,Agliardi:2022ghn,deLejarza:2023qxk}, including their application to loop Feynman integrals and decay rates at next-to-leading order~\cite{deLejarza:2024pgk,deLejarza:2024scm}. In this work, we delve into the quantum determination of the causal structure of multiloop Feynman diagrams previously analysed in ~\cite{Ramirez-Uribe:2021ubp,Clemente:2022nll}.

Improving the accuracy of theoretical predictions for high-energy colliders is crucial because of the high demands this field will face~\cite{Strategy:2019vxc}, in particular during the current Run 3 of the CERN's Large Hadron Collider (LHC), the planned high-luminosity LHC stage~\cite{Gianotti:2002xx}, and future collider projects~\cite{Abada:2019lih,Djouadi:2007ik,Roloff:2018dqu,CEPCStudyGroup:2018ghi}. As far as perturbative Quantum Field Theory is concerned, one of the main needs is the study and appropriate treatment of the quantum fluctuations associated to multiloop Feynman diagrams. In accordance with this need, there has been a remarkable effort in the field to provide powerful frameworks to meet this challenge~\cite{Heinrich:2020ybq}.

Aiming for improving the efficiency of multiloop-level computations with a proper treatment of singularities, we work in the framework of the Loop-Tree Duality (LTD)~\cite{Catani:2008xa,Bierenbaum:2010cy,Bierenbaum:2012th,Buchta:2014dfa,Buchta:2015wna,Hernandez-Pinto:2015ysa,Sborlini:2016gbr,Sborlini:2016hat,Tomboulis:2017rvd,Driencourt-Mangin:2017gop,Jurado:2017xut,Driencourt-Mangin:2019aix,Runkel:2019yrs,Aguilera-Verdugo:2019kbz,Runkel:2019zbm,Capatti:2019ypt,Driencourt-Mangin:2019yhu,Capatti:2019edf,Plenter:2019jyj,Plenter:2020lop,Prisco:2020kyb}. In particular, the latest achievements in LTD are based on its most distinctive feature: the existence of a manifestly causal representation and its connection with directed acyclic graph (DAG) configurations in graph theory~\cite{Verdugo:2020kzh,snowmass2020,Aguilera-Verdugo:2020kzc,Aguilera-Verdugo:2020nrp,Ramirez-Uribe:2020hes,Sborlini:2021owe,TorresBobadilla:2021ivx,TorresBobadilla:2021dkq,Aguilera-Verdugo:2021nrn, Rios-Sanchez:2024xtv,Ramirez-Uribe:2024rjg,LTD:2024yrb}. A comprehensive review of classical algorithms in graph theory is presented in Ref.~\cite{Even20111}.

The emergence of noncausal singularities in the customary Feynman representation of loop integrals implies considerable numerical instabilities, whereas in the LTD representation, noncausal singularities are explicitly absent, providing the advantage of significantly more stable integrands~\cite{Aguilera-Verdugo:2020kzc,Ramirez-Uribe:2020hes}.
The fact that a Feynman propagator has two on-shell states, which are therefore naturally encoded in a qubit, makes quantum computing a potentially efficient way of dealing with some aspects of multiloop Feynman integrals.
In Ref.~\cite{Ramirez-Uribe:2021ubp}, an amplitude amplification or Grover's based~\cite{Grover:1997fa,Boyer:1996zf} quantum algorithm was proposed to unfold the causal configurations of multiloop Feynman diagrams. A similar study based on a variational quantum eigensolver~(VQE) approach was presented in Ref.~\cite{Clemente:2022nll}. The quantum algorithm of Ref.~\cite{Ramirez-Uribe:2021ubp} is quite expensive in terms of quantum resources, while the VQE approach~\cite{Clemente:2022nll} requires fewer quantum resources but longer runs. 

In this paper, we present an enhanced quantum algorithm for querying causality of multiloop Feynman diagrams with a significant reduction in qubit requirements and quantum circuit area through a more systematic and efficient logic implementation for identifying directed acyclic graphs~\cite{EPpatent}. The improvement achieved allows us to explore multiloop topologies of higher complexity that were previously inaccessible due to the current limitations of quantum simulators.

\section{Causality from the loop-tree duality}
\label{sec:LTD}

Generic loop integrals and scattering amplitudes, with $n$ propagators and $P$ external legs are defined in the Feynman representation as integrals in the Minkowski space of~$L$ loop momenta
\begin{equation}\label{eq:FeynmanRep}
    \mathcal{A}_{\text{F}}^{(L)} 
    = \int_{\ell_1 \ldots \ell_L} {\cal N} \big(\{\ell_s\}_L, \{p_j\}_P \big) 
    \prod_{i=1}^n G_{\text{F}}(q_i)~,
\end{equation}
where the momenta of the Feynman propagators, $q_i^\mu$, are linear combinations of the loop and external momenta, and the numerator ${\cal N}$ is given by the specific theory. The integration measure in dimensional regularisation~\cite{Bollini:1972ui,tHooft:1972tcz} reads $\int_{\ell_s} = -\imath \mu^{4-d} \int {\rm d}^d \ell_s/(2\pi)^d$, with $d$ the number of space-time dimensions and $\mu$ an arbitrary energy scale. Feynman propagators are conveniently written as
\beq
    G_{\text{F}}(q_i) = \frac{1}{\big(q_{i,0}-\qon{i}\big)\big(q_{i,0}+\qon{i}\big)}~,
\label{eq:propagator}
\eeq
where $\qon{i} = \sqrt{\qb_i^2+m_i^2-\ii}$ are the on-shell energies, with $q_{i,0}$ and $\qb_i$ the temporal and spacial components of $q_i$, respectively, and $m_i$ the mass of the propagating particle. The infinitesimal factor $\ii$ is the usual complex prescription of a Feynman propagator.

Singularities in the integrand of~\Eq{eq:FeynmanRep} arise when Feynman propagators are set on shell, explicitly, when the energy component $q_{i,0}$ takes one of the values $\pm\qon{i}$ in \Eq{eq:propagator}. An important fact to take into account is that not all potential singular configurations of the integrand in~\Eq{eq:FeynmanRep} lead to physical singularities of the integral. In the case of the Feynman representation of loop integrals, noncausal singularities are unavoidable. Regarding the LTD framework, its more relevant feature is the existence of a manifestly causal representation, where noncausal singularities are absent. The direct LTD representation of~\Eq{eq:FeynmanRep} is computed by applying the Cauchy's residue theorem through the evaluation of nested residues~\cite{Verdugo:2020kzh}. To obtain the LTD causal representation we sum over all the nested residues, explicitly cancelling all the noncausal contributions. Furthermore, the LTD causal representation is found to be reinterpreted in terms of entangled causal thresholds~\cite{Aguilera-Verdugo:2020kzc,Ramirez-Uribe:2020hes}. The analytical causal reconstruction is achieved by matching all combinations of $n-L$ threshold configurations that are causally compatible to each other, leading to the LTD causal representation, 
\beq
    \mathcal{A}_{\text{D}}^{(L)} = \int_{\boldsymbol{\ell}_1 \ldots \boldsymbol{\ell}_L} 
    \frac{1}{x_n} \sum_{\sigma  \in \Sigma} \frac{{\cal N}_{\sigma(i_1, \ldots, i_{n-L})}}{\lambda_{\sigma(i_1)}^{h_{\sigma(i_1)}} \cdots \lambda_{\sigma(i_{n-L})}^{h_{\sigma(i_{n-L})}}}
    + (\lambda_p^+ \leftrightarrow \lambda_p^-)~,
\label{eq:CausalRep}
\eeq
with $x_n = \prod_{i=1}^n 2\qon{i}$ and $h_{\sigma(i)}=\pm$.
The Feynman propagators from \Eq{eq:FeynmanRep} are substituted in Eq.~\eqref{eq:CausalRep} by causal propagators of the form $1/\lambda_p^\pm$, with
\beq
    \lambda_p^\pm = \sum_{i\in p} \qon{i} \pm k_{p,0}~,
\eeq
where $p$ is a partition of the on-shell energies, and $k_{p,0}$ is a linear combination of the external-momenta energy components. Given the sign of $k_{p,0}$, either $\lambda_p^-$ or $\lambda_p^+$ becomes singular after all propagators in the partition $p$ are set on shell. The combinations of entangled causal propagators represent causal threshold configurations that can occur simultaneously which are collected in the set $\Sigma$. 

A crucial step in the procedure for obtaining the causal representation shown in Eq.~\eqref{eq:CausalRep}, is the identification of diagram configurations admitting compatibility among causal thresholds (causal singular configurations), which is seen as querying multiple solutions over an unstructured database~\cite{Grover:1997fa,Boyer:1996zf}.
This exercise is equivalent to select all the DAG configurations of an undirected graph in graph theory. In the following section we present a systematic and efficient quantum algorithm to address the identification of causal configurations of multiloop Feynman diagrams.

\section{Quantum causal querying}
\label{sec:QuantumGrover}

The proposed algorithm recovers the central ideas from Ref.~\cite{Ramirez-Uribe:2021ubp} on amplitude amplification or Grover's quantum algorithm and considers a generalised description of multiloop topologies by treating them as directed graphs made of edges and eloops~\cite{TorresBobadilla:2021ivx,Sborlini:2021owe}. An edge is defined as the union of multiple Feynman propagators connecting two interaction vertices. Once propagators are substituted by edges, the remaining loops of the graph define the number of eloops, which is smaller than the number of loops of the original Feynman diagram. The convenience of working with eloops in a causality scenario is given by the fact that the only feasible causal configurations are those for which the momentum flows of all propagators belonging to each edge are aligned in the same direction.

The distinguishing idea introduced in this paper is a change of logic in the construction of the oracle operator. The tagging of the unidirectional cycles representing eloops is done through the application of multicontrolled Toffoli (MCX) gates and NOT (Pauli-$X$) gates, having as a core difference with Ref.~\cite{Ramirez-Uribe:2021ubp} the absence of binary clauses comparing adjacent edges. This logical procedure significantly simplifies the design and encoding of the corresponding quantum circuits. To evaluate the efficiency of the proposal, we focus on the number of qubits required for the algorithm implementation and on the quantum circuit depth circuit area. Before describing the algorithm, we review some relevant concepts for a better understanding of the work developed. 

\begin{figure}[tb!]
\centering
\includegraphics[scale=0.8]{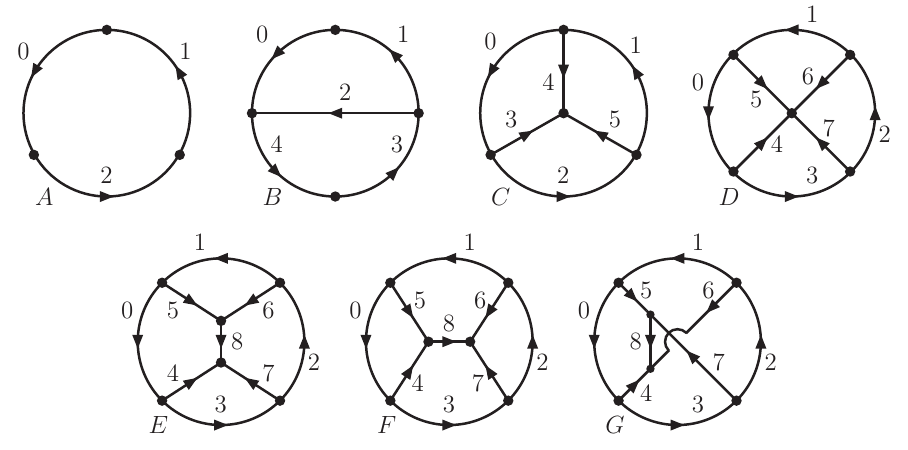}
\caption{From left to right and top to bottom: one eloop with three vertices, two eloops with four vertices, three eloops with four vertices, four eloop with a four-point contact interaction, and four eloops $t$-, $s$- and $u$-channels.  The direction of the arrows corresponds to the $\ket{1}$ states. External momenta are attached to the interaction vertices. 
\label{fig:Topologies1}}
\end{figure}

Regarding the efficiency of the amplitude amplification, it is important to note the relevance of the mixing angle, defined as $\theta = \arcsin \sqrt{r/N}$, that measures the ratio of the~$r$ states to be queried over the total number of possible states, $N=2^n$, and the optimal number $t$ of the algorithm iterations. If $r \sim N/4$, then $\theta \sim \pi/6$, and one iteration is sufficient because $\sin^2 \theta_t \sim 1$ with $\theta_t = (2t +1) \, \theta$. It has been shown from classical~\cite{Aguilera-Verdugo:2020kzc,Ramirez-Uribe:2020hes} and quantum~\cite{Ramirez-Uribe:2021ubp} analysis of the multiloop topologies depicted in Fig.~\ref{fig:Topologies1} that the number of causal configurations is typically about one half of the total number of possible states. This implies $\theta \sim \pi/4$, which does not lead to any reasonable amplitude amplification. However, given a causal solution of a loop Feynman diagram, the mirror state with all the momentum flows reversed is also a causal solution. Therefore, for one of the qubits representing an edge, we tag only one of its states  so that the number of states to be queried is $r\sim N/4$, which is the optimal value for amplitude amplification in a single iteration. The total set of causal solutions is then obtained by reversing the momentum flows of all the causal states queried.
In the case that halving the number of solutions is not enough to adjust the ratio~$r/N$ in the required range, then we introduce an ancillary qubit~\cite{Nielsen:2012yss} to increase the number of total states without increasing the number of solutions.

The implementation of the quantum algorithm requires three registers, $\ket{e}$, $\ket{a}$ and $\ket{\text{out}}$.
The first register, $\ket{e}$, encodes the states of the edges. For each edge, $\ket{0}$ represents propagation in one direction, and $\ket{1}$ in the opposite direction. We define the Boolean functions
\beq\label{eq:setedges} 
    s_j \equiv \underset{i\in s_j}{\bigwedge} e_{i} ~, \qquad
    \bar s_j \equiv \underset{i\in s_j}{\bigwedge} \bar e_{i} ~,
\eeq
where $s_j$ labels the edge set $e_i$ depending on the same linear combination of loop momenta. The bar in $\bar{s}_j$ denotes that the qubit states in this set are inverted with NOT gates, $\ket{\bar e_{i}} = X\ket{e_{i}}$. The number of edges in the set $s_j$ is denoted by $n_{s_j}$ and the total number of edges describing the corresponding multiloop topology is $n=\sum_j n_{s_j}$, which is the size of the register $\ket{e}$. The second required register is $\ket{a}$. It collects the causal conditions in eloop clauses, and its size is $m$, corresponding to the number of sub-eloops that need to be tested. The last register, $\ket{\text{out}}$, stores the validation of the causal solutions in what is known as the oracle's marker. 
The algorithm takes as a first step the initialisation of all the quantum registers. The register encoding the edges' states, $\ket{e}$, is initialised in a uniform superposition by applying Hadamard gates to each of its $n$ qubits, $\ket{e}=H^{\otimes n}\ket{0}$. For the eloop clauses, the $m$ qubits in the register $\ket{a}$ are initialised to $\ket{1}$ by applying NOT gates, $\ket{a}= X^{\otimes m} \ket{0}$. The oracle's marker qubit, $\ket{\text{out}}$, is initialised to one of the Bell states by applying $\ket{\text{out}}=H(X\ket{0})=H\ket{1}\equiv\ket{-}$.

Following to the construction of the oracle operator, it is important to recall that causal configurations are identified as DAG configurations, therefore, the eloop clauses are established to suppress directed cyclic configurations. The eloop clauses are implemented through the application of multicontrolled Toffoli gates and NOT gates that validate whether a configuration is cyclic or acyclic. To provide a convenient notation, we define the multicontrolled Toffoli operation based on Boolean clauses by
\begin{align}\label{eq:toffDef}
    \toff(S_j) = \underset{k \in S_j}{\bigwedge}s_k ~, \qquad 
    \bar{\toff}(S_j) = \underset{k \in S_j}{\bigwedge}\bar{s}_k ~,
\end{align}
and
\begin{align}\label{eq:ToffDef}
    \Toff(S_j)&= \toff(S_j) \bigvee \bar{\toff}(S_j) ~,
\end{align}
where the capital $S_j$ contains the edge set generating the cyclic configuration involving the sub-eloop $j$. Both conditions in~\Eq{eq:toffDef} select the same sub-eloop but with opposite edge directions. The composite condition in \Eq{eq:ToffDef} embraces both directions of the given sub-eloop associated to $S_j$. Let us recall that the eloop clause register $\ket{a}$ is initialised to~$\ket{1}$, therefore, applying one of the conditions in Eqs.~(\ref{eq:toffDef}) and (\ref{eq:ToffDef}) unmarks a noncausal configuration from $\ket{a}$. 

An important aspect to consider relating eloop clauses is whether or not the set $S_j$ involves a fixed qubit, that is, a qubit tagged in a definite state $\ket{0}$ or $\ket{1}$. If the eloop clause condition does not involve a fixed qubit, we apply \Eq{eq:ToffDef}, which considers two multicontrolled Toffoli gates to validate both directions. The validation of both directions is stored in a single qubit $\ket{a_i}$, due to the fact that both configurations are mutually exclusive, i.e., applying one multi-controlled Toffoli gate does not overlap over the second one. In the case where the eloop clause involves a fixed qubit, the condition to probe involves only one direction, explicitly, only one of the two Boolean conditions given in \Eq{eq:toffDef} is tested.

\begin{figure}[tb!]
\centering
\includegraphics[scale=2.5]{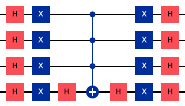}
\caption{Diffusion operator for four qubits.
\label{fig:diffusionoperator}}
\end{figure}

After the initialisation of the registers and setting all the eloop clauses, the oracle operator is given by 
\begin{equation}
    U_w \ket{e} \ket{a} \ket{\text{out}} = 
    (-1)^{f(a,e)} \ket{e} \ket{a} \ket{\text{out}}~,
    \label{eq:oracle}
\end{equation}
where $f(a,e)=\left( \bigwedge_i a_i \right) \wedge e_0$, and $e_0$ is the edge whose state is fixed. The procedure continues by performing the oracle operations in inverse order to restore every qubit but $\ket{\text{out}}$ to its initial state. The oracle operator tags the causal configurations according to~\Eq{eq:oracle} but does not alter the probabilities of these states. Then, a diffusion operator, $U_s$, amplifies the amplitude of the tagged states, increasing their probability in a measurement of the qubits. In particular, if $\ket{w}$ represents a uniform superposition of the tagged or winning states, and $\ket{e_\perp}$ is the orthogonal state, then
\beq \label{eq:diffusion_operator}
U_s\left( U_w\ket{e}\right) = \cos(3\theta)\ket{e_{\perp}} +\sin(3\theta)\ket{w} ~.
\eeq
The probabilities of the orthogonal and winning states are $\cos^2(3\theta)$ and $\sin^2(3\theta)$, respectively.  If $\sin^2(3\theta) \sim 1$, the probability of the orthogonal states is very suppressed and the measurement of the qubits in the register $\ket{e}$ selects the desired states with higher probabilities. The diffusion operator implemented in the quantum circuit is shown in Fig.~\ref{fig:diffusionoperator} and is based on the diffusion operator described in the IBM Qiskit website \footnote{\texttt{\href{http://qiskit.org/}{http://qiskit.org/}}}. 

The quantum algorithm presented in this paper has the advantage that the quantum circuit depth depends just on the number of eloop clauses and not on the number of edges per set, whereas the quantum circuit depth of the quantum algorithm using binary clauses~\cite{Ramirez-Uribe:2021ubp} increases when additional edges per set are considered.

Let us recall that the quantum circuit depth~\cite{Depth:5769} is defined as the integer number of gates that need to be executed along the longest path of the circuit from input to output, moving forward in time along qubit wires. The input is understood as the preparation of the qubits and the output as the measurement gate. The quantum circuit depth is computed through the Qiskit routine, \texttt{QuantumCircuit.depth()}~\footnote{\href{https://qiskit-test.readthedocs.io/en/latest/api/qiskit.circuit.QuantumCircuit.html}{\texttt{https://qiskit-test.readthedocs.io}}}. An important feature of the quantum circuit depth is the impact of quantum gates acting on no common qubits, allowing to perform them at the same time step. On the other hand, if the gates act on at least one common qubit they have to be applied in different time-steps, increasing by one unit the quantum circuit depth. 

The output of the quantum algorithm after a measurement is the selection of a single configuration. In order to identify all the solutions it is required to prepare and measure a certain number of times (shoots), having as an output a frequency histogram. An estimation of the minimal number of shots required to distinguish causal from noncausal configurations with a statistical significance of $\Delta \sigma$ standard deviations in a quantum simulator~\cite{Ramirez-Uribe:2021ubp} is given by 
\beqn
N_{\rm shots} \approx r \left(\Delta \sigma\right)^2 \left(1 + {\cal O}(\cos^2(\theta_t)) \right) \, ,
\label{eq:shots}
\eeqn
assuming an efficient amplification of the causal states, i.e. $\cos(\theta_t) \sim 0$.

\begin{figure}[tb!]
    \centering
    \raisebox{35pt}{
    \includegraphics[scale=0.2]{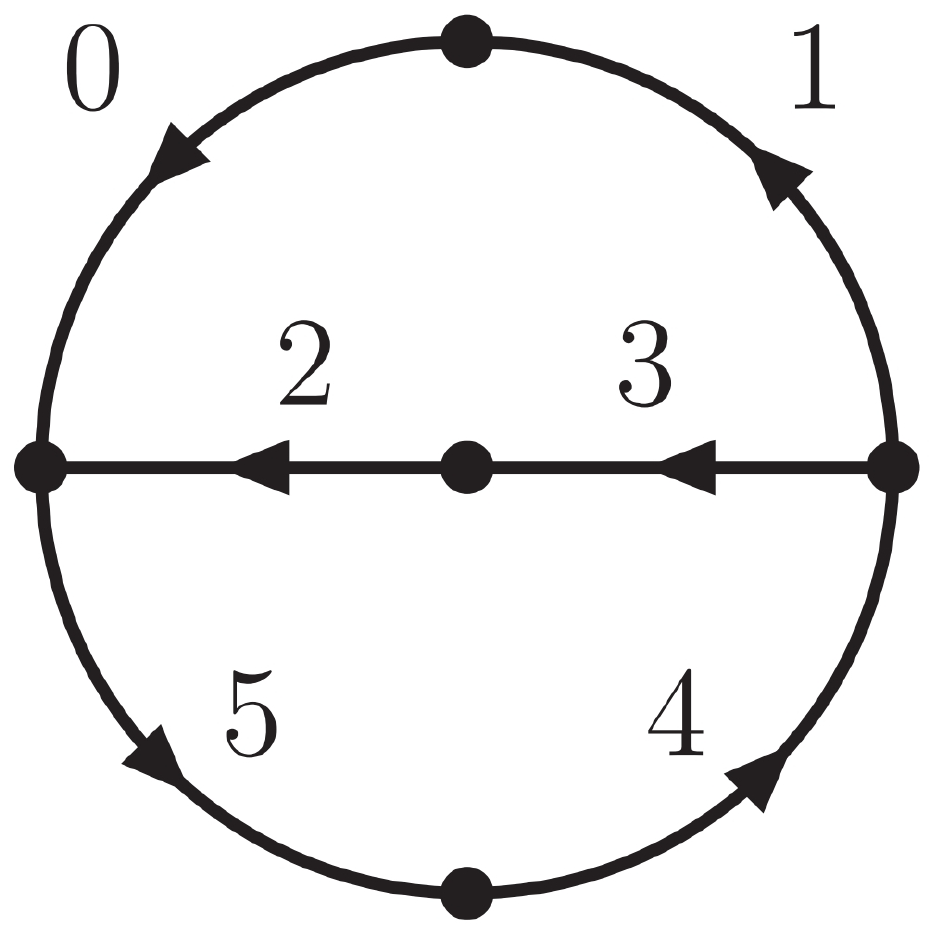}}\qquad\qquad
    \includegraphics[scale=0.33]{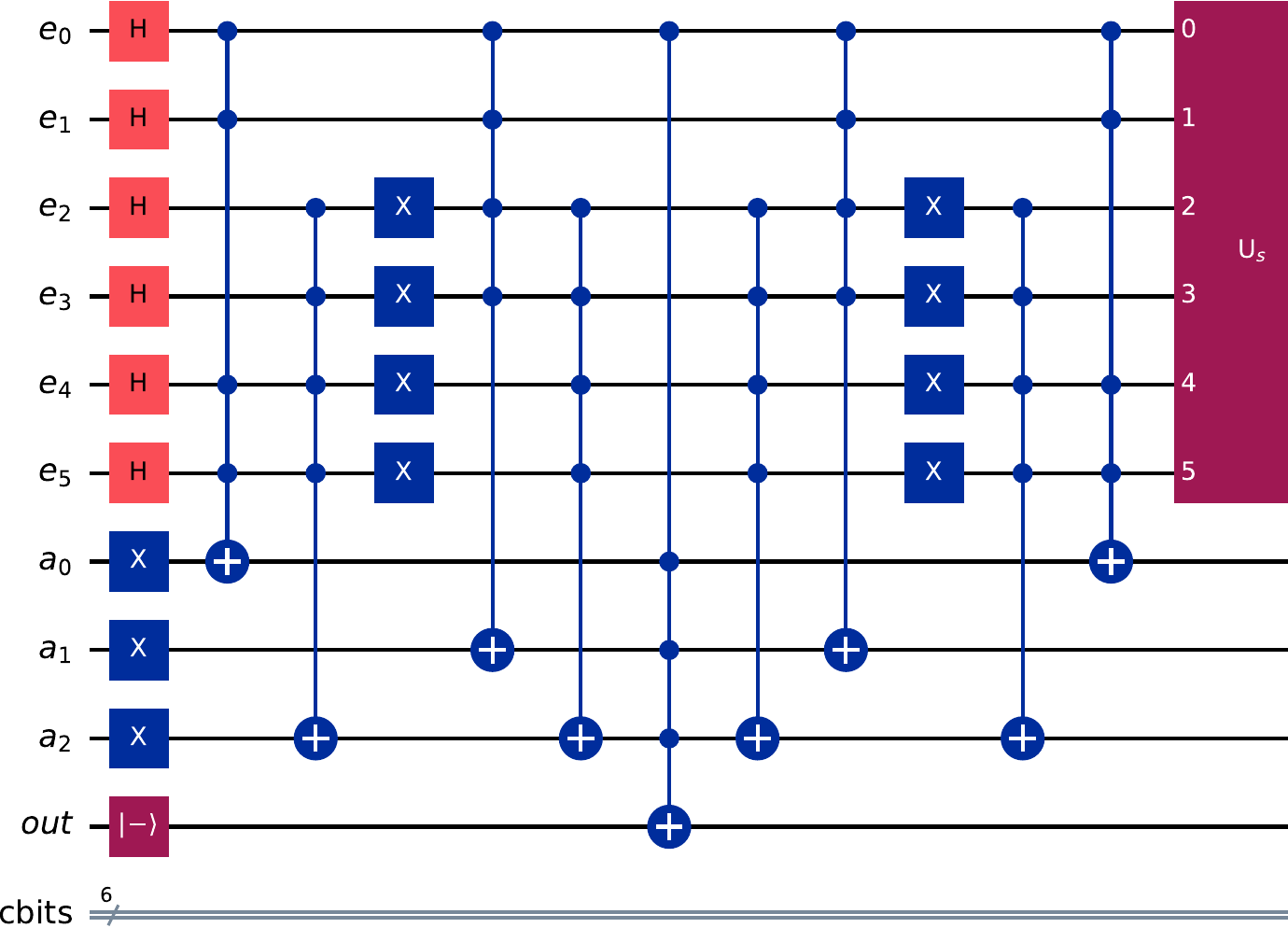}
\includegraphics[width=0.84\textwidth]{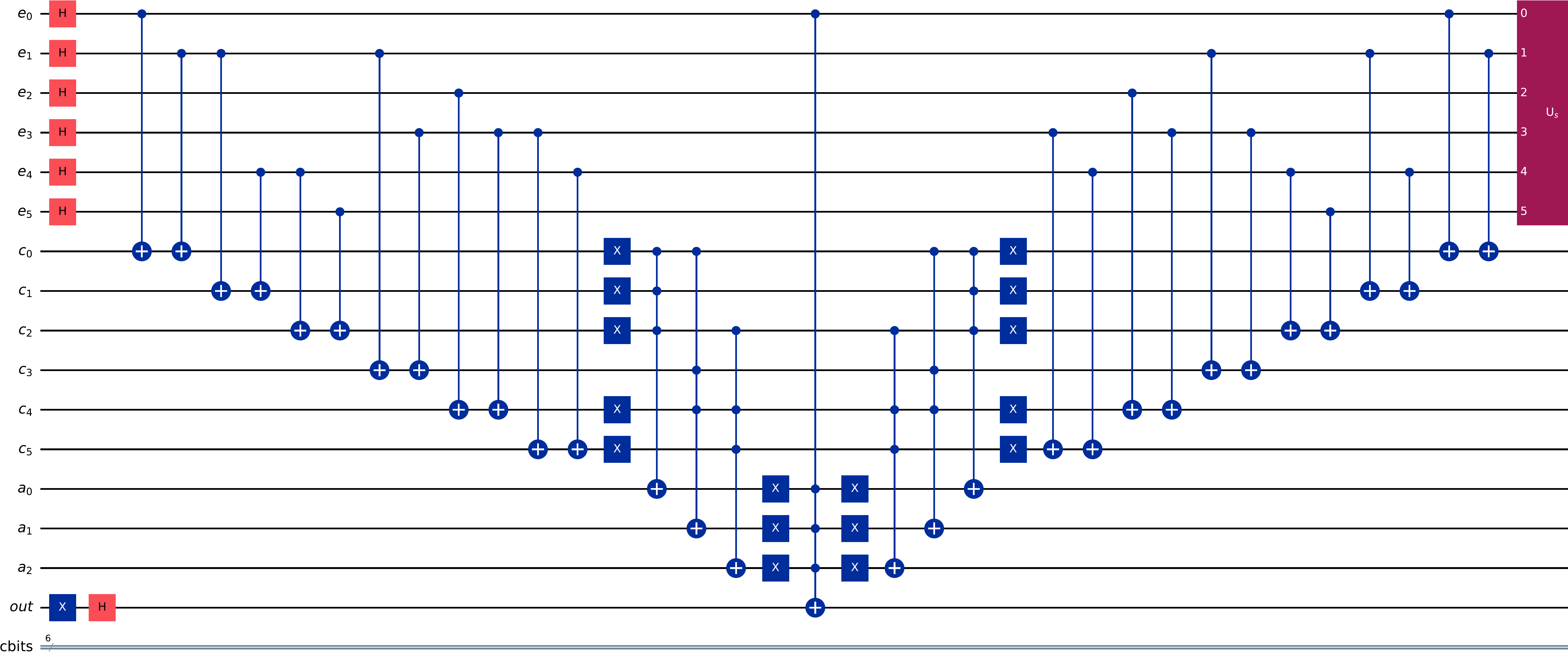}
    \caption{Two-eloop topology with two edges per edge set (top left). Quantum circuit with an oracle operator based on multicontrolled Toffoli gates and 
    NOT gates (top right). Implementation of the quantum algorithm presented in Ref~\cite{Ramirez-Uribe:2021ubp} using CNOT gates for the binary clauses and  multicontrolled Toffoli gates for the registers $\ket{a}$ and $\ket{\text{out}}$ (bottom).
    \label{fig:2eloop2EdgesOracle}}
\end{figure}
\begin{figure}[tb!]
    \centering
    \includegraphics[width=\textwidth]{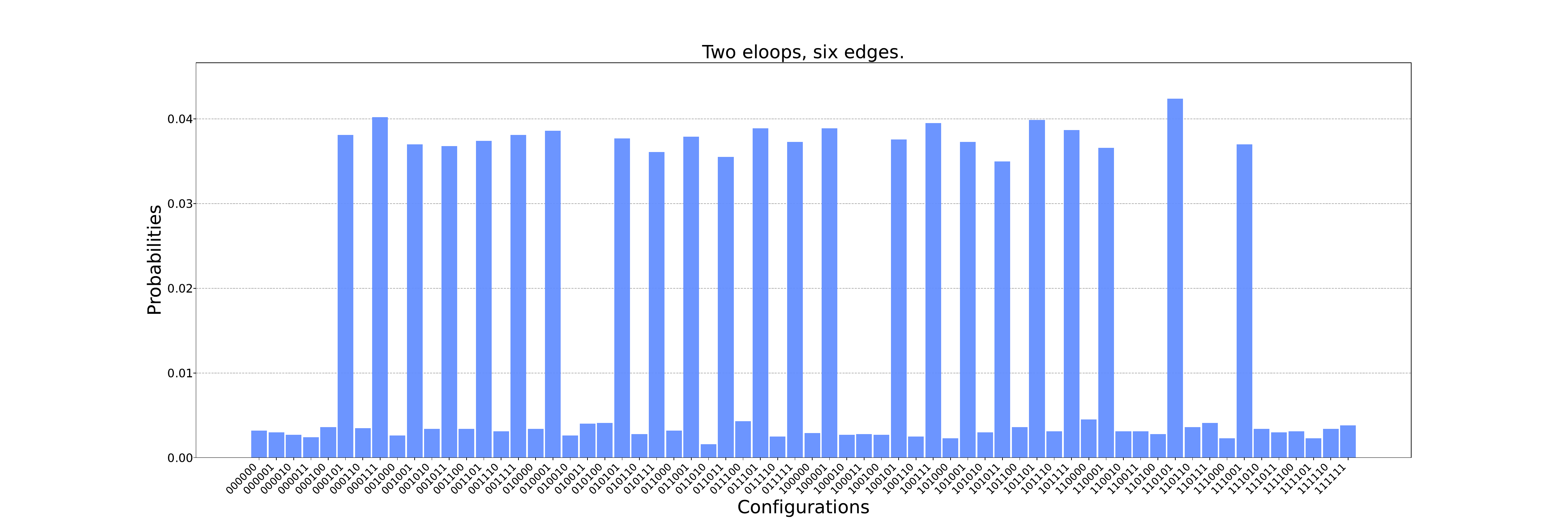}
    \caption{Output for the algorithm implementation of the two-eloop topology with six edges. The total number of causal configurations is 23 out of 64.}
    \label{fig:2eloopHistogram}
\end{figure}

To illustrate the algorithm implementation, let us start with the two-eloop topology with three sets of edges depicted in the top left of Fig.~\ref{fig:2eloop2EdgesOracle}. We consider two edges per set to include three additional vertices representing external particles.
Concretely, we have $n=6$, $s_0 = 0\wedge 1$, $s_1 = 2 \wedge 3$ and $s_2 = 4 \wedge 5$, using the short-hand notation $i\equiv e_i$. As previously stated, we start by preparing the three registers: the $\ket{e}$ register in a uniform superposition $H^{\otimes n}\ket{e_i}$, the register $\ket{a}$ initialised in the state $\ket{1}$ and the oracle's marker in the state $\ket{-}$, as shown in the first time step of the quantum circuit of Fig.~\ref{fig:2eloop2EdgesOracle}~(top right). The eloop clauses are given by
\begin{align}\label{eq:clauses2eloop}
    a_0^{(2)} &= \toff(s_0\wedge s_2) ~,\nn\\
    a_1^{(2)} &= \toff(s_0\wedge \bar{s}_1) ~,\nn\\
    a_2^{(2)} &= \Toff(s_1\wedge s_2) ~,
\end{align}
where $S_0=s_0\wedge s_2$, $S_1=s_0\wedge \bar{s}_1$ and $S_2=s_1\wedge s_2$. The eloops clauses in~\Eq{eq:clauses2eloop} are implemented by multicontrolled Toffoli gates taking as control the qubits in the set $S_j$ and as target the qubit $\ket{a_j}$ with $j=0,1,2$. Explicitly, $a_0^{(2)}$ and $a_1^{(2)}$ are implemented by a single multicontrolled Toffoli gate, whereas $a_2^{(2)}$ is implemented by two multicontrolled Toffoli gates.

For the oracle's marker, another multicontrolled Toffoli gate is used. The initial mixing angle is $\theta=36.8^\circ$ (see Table.~\ref{tab:comparative_2}) by fixing the qubit $\ket{e_0}$. The multicontrolled Toffoli gate implementing the oracle's marker takes as control the qubits in the register $\ket{a}$ and the qubit $\ket{e_0}$, and as target the qubit $\ket{\text{out}}$. After applying the oracle operations in reverse order, we apply the diffusion operator to the register~$\ket{e}$. Table.~\ref{tab:comparative_2} shows the results of the two quantum circuits in Fig.~\ref{fig:2eloop2EdgesOracle}, going from $16$ to $10$ total number of qubits for implementation, and from $22$ to $14$ quantum circuit depth.

In the following sections we explore in detail the performance of the proposed quantum algorithm by analysing the number of qubits needed for the implementation and the corresponding theoretical quantum circuit depth.

\begin{table}
\begin{center}
\begin{tabular}{llccccrr} \toprule \small
\bfseries  Fig. & \bfseries eloops (edges) & 
Qubits & $\substack{\text{Quantum} \\ \text{Depth}}$& 
 $ \theta$ & $\sin^2(\theta_t)$ & $\substack{\text{Causal} \\ \text{states}}$ &
 $\substack{\text{Total} \\ \text{states}}$\\ \midrule 
\ref{fig:2eloop2EdgesOracle} & two ($6$)  & $\ph{0}10~|~16$ & $14~|~22$ & $36.8^\circ$ &  $0.87$ & $23$ & $64$ \\
\bottomrule 
\end{tabular}
\end{center}
\caption{Quantum circuit requirements of the quantum algorithms used to analyse the two-eloop topology in Fig.~\ref{fig:2eloop2EdgesOracle}. The first number in the second and third columns is from the quantum algorithm proposed in this work, whereas the second number corresponds to the binary-clause algorithm~\cite{Ramirez-Uribe:2021ubp}. The columns with a single entry have the same value for both algorithms. The numbers in the last two columns are the number of causal configurations and the total number of states, respectively. \label{tab:comparative_2}
}
\end{table} 

\section{Efficiency assessment}
\label{sec:Comparative}

In this section we assess the efficiency of the proposed quantum algorithm by comparing the implementation results with the quantum algorithm and the multiloop topologies analysed in Ref.~\cite{Ramirez-Uribe:2021ubp}. These multiloop topologies are depicted in Fig.~\ref{fig:Topologies1} considering one edge in each edge set $s_i$. The results are summarised in Table~\ref{tab:comparative} which displays the number of qubits required for the implementation, the theoretical quantum circuit depth, the related mixing angle and the number of causal configurations.

\begin{table}
\begin{center}
\begin{tabular}{llccccrr} \toprule \small
\bfseries  Fig. & \bfseries eloops (edges) & 
Qubits & $\substack{\text{Quantum} \\ \text{Depth}}$& 
 $ \theta$ & $\sin^2(\theta_t)$ & $\substack{\text{Causal} \\ \text{states}}$ &
 $\substack{\text{Total} \\ \text{states}}$\\ \midrule 
\ref{fig:Topologies1}A & one ($3$)      & $\ph{0}5~|~\ph{0}7$ & $\ph{0}6~|~18$ & $37.7^\circ$ & $0.84$ & $3$ & $8$ \\ 
\ref{fig:Topologies1}B & two ($5$)      & $\ph{0}9~|~14$ & $12~|~21$ & $32.0^\circ$ &  $0.99$ & $9$ & $32$ \\
\ref{fig:Topologies1}C & three ($6$)    & $11~|~19$ & $18~|~24 $ & $25.7^\circ$ &  $0.95$ & $12$ & $64$ \\
\ref{fig:Topologies1}D & four$^{(c)}$ ($8$) & $14~|~25$ & $16~|~23$ & $33.5^\circ$ & $0.97$ & $39$ & $256$ \\
\ref{fig:Topologies1}E, \ref{fig:Topologies1}F & four$^{(t,s)}$ ($9$) & $15~|~28$ & $20~|~25$ & $26.5^\circ$ & $0.97$ & $102$ & $512$ \\
\ref{fig:Topologies1}G & four$^{(u)}$ ($9$) & $19~|~33$ & $32~|~28$ & $28.3^\circ$ & $0.99$ & $115$ & $512$\\
\bottomrule 
\end{tabular}
\end{center}
\caption{Quantum resources required and quantum circuit depth of the quantum algorithms used to analyse the multiloop topologies in Fig.~\ref{fig:Topologies1}. The first number in the third and fourth columns is from the quantum algorithm proposed in this work, whereas the second number corresponds to Ref.~\cite{Ramirez-Uribe:2021ubp}. The columns with a single entry have the same value for both algorithms. The numbers in the last two columns are the number of causal configurations and the total number of states, respectively.}\label{tab:comparative}
\end{table}

The columns of total number of qubits and quantum circuit depth have two numbers: the first one gives the data on the current approach and, the second one the information of the quantum algorithm using binary clauses. The remaining columns are: the initial mixing angle, an indicator for the appropriate selection of the iteration number and the number of causal states. From Table~\ref{tab:comparative}, we observe that in all cases the initial mixing angle is close to $\pi/6$, suggesting the application of only one iteration; from the fifth column we have that the value of the mixing angle after the first iteration approaches $\pi/2$, leading to the optimal probability amplification, $\sin^2(\theta_t)\sim 1$ with $t=1$.

Counting the number of qubits needed, the proposed quantum algorithm lowers significantly the total number of qubits required for the implementation with respect to a binary-clause algorithm, going from a reduction of two qubits in the case of one eloop with three vertices, to a reduction of fourteen qubits in the $u$-channel at four eloops. Regarding the theoretical quantum circuit depth, we notice that the new approach yields smaller values for all multiloop topologies, with the only exception of the four-eloop $u$-channel, which is a nonplanar diagram. It is important to note that the values shown in Table~\ref{tab:comparative} correspond to multiloop configurations with one edge per edge set. For these configurations it is sufficient to satisfy the Boolean clauses in \Eq{eq:toffDef} for sub-eloops involving a small number of edge sets, while longer strings of edge sets need to be checked when multiple edges in the edge sets are considered. However, the number of eloop clauses reaches a maximum that depends on the multiloop topology and does not scale with the number of edges (see Table~\ref{tb:newtopogral}), while the number of binary clauses scales with the number of edges. In Section~\ref{sec:Application}, we study multiloop topologies with multiple edges in each edge set, which require testing the maximum number of eloop clauses.\\

\begin{figure}
\centering
\includegraphics[width=\textwidth]{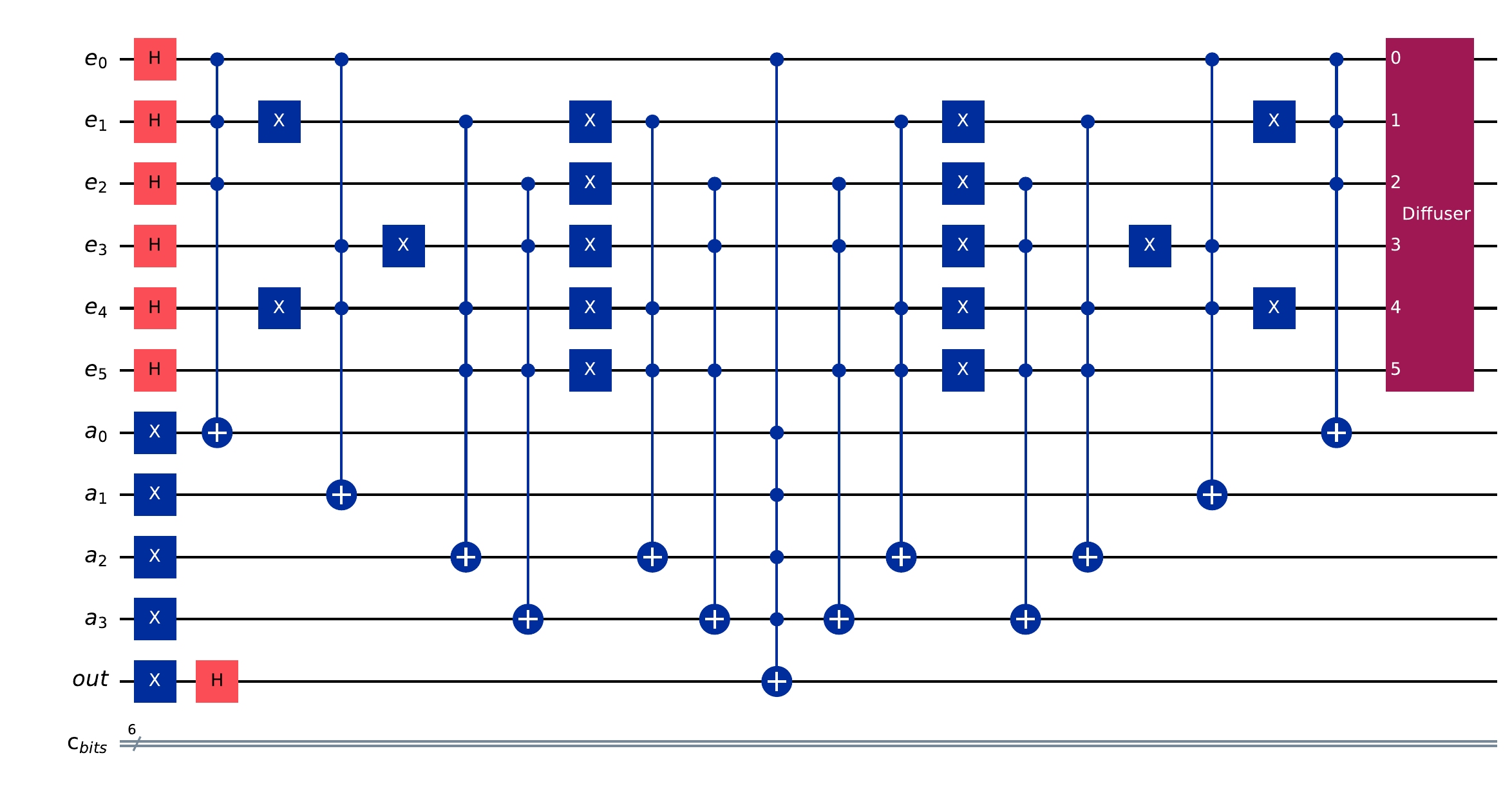}
\caption{Quantum circuit for the three-eloop topology with six edges (Fig.~\ref{fig:Topologies1}C). The output is 12 causal configurations out of 64 states.
\label{fig:qc3eloops6}}
\end{figure}
Continuing with the analysis of the quantum circuit depth, we observe in Table~\ref{tab:comparative} that the quantum circuit depth for the three-eloop diagram (Fig.~\ref{fig:Topologies1}C) is greater than the quantum circuit depth for the four-eloop diagram with one four-particle interaction vertex (Fig.~\ref{fig:Topologies1}D, also called N$^3$MLT in Ref.~\cite{Ramirez-Uribe:2020hes}). To better understand this behaviour, we compare the eloop clauses needed to implement the two topologies, and analyse the presence of multicontrolled Toffoli gates acting on noncommon qubits along the circuits. The eloop clauses required for the three-eloop topology are
\begin{align}\label{eq:clauses3}
    a_0^{(3)} &= \toff (s_0 \wedge s_1 \wedge s_2) ~, &
    a_1^{(3)} &= \toff (s_0 \wedge s_3 \wedge \bar{s}_4) ~, \nn\\[5pt]
    a_2^{(3)} &= \Toff (s_1 \wedge s_4 \wedge \bar{s}_5) ~, &
    a_3^{(3)} &= \Toff (s_2 \wedge \bar{s}_3 \wedge s_5) ~, 
\end{align}
and for the four-eloop topology are
\begin{align}\label{eq:clauses4}
    a_0^{(4)} &= \toff (s_0 \wedge s_1 \wedge s_2 \wedge s_3) ~, &
    a_1^{(4)} &= \Toff (s_2 \wedge s_6 \wedge \bar{s}_7) ~, \nn\\
    a_2^{(4)} &= \toff (s_0 \wedge s_4 \wedge \bar{s}_5) ~, &
    a_3^{(4)} &= \Toff (\bar{s}_1 \wedge \bar{s}_5 \wedge s_6) ~, \nn\\
    a_4^{(4)} &= \Toff (\bar{s}_3 \wedge s_4 \wedge \bar{s}_7) ~. 
\end{align}
In the three-eloop topology, Fig.~\ref{fig:Topologies1}C, the eloop clauses in \Eq{eq:clauses3} are implemented in the quantum circuit using six multicontrolled Toffoli gates. The clauses $a_0^{(3)}$ and $a_1^{(3)}$ validate only one direction of the corresponding sub-eloops, requiring a single multicontrolled Toffoli gate, while $a_2^{(3)}$ and $a_3^{(3)}$ test the sub-eloops in both directions, requiring two multicontrolled Toffoli gates each. In adition, the oracle operator needs seven NOT gates. In this case, given that the Toffoli gates act in at least one common edge, they must be executed in different time-steps to avoid interference. Note that the final number of multicontrolled Toffoli gates and $X$ gates is doubled to return the $\ket{a}$ register to its initial state, and an extra multicontrolled Toffoli gate is required that includes the oracle marker. The initialization of the $\ket{a}$ and $\ket{\text{out}}$ registers also needs additional NOT gates. The quantum circuit for the three-eloop topology~\ref{fig:Topologies1}C is shown in Fig.~\ref{fig:qc3eloops6}.

\begin{figure}
\centering
\includegraphics[width=\textwidth]{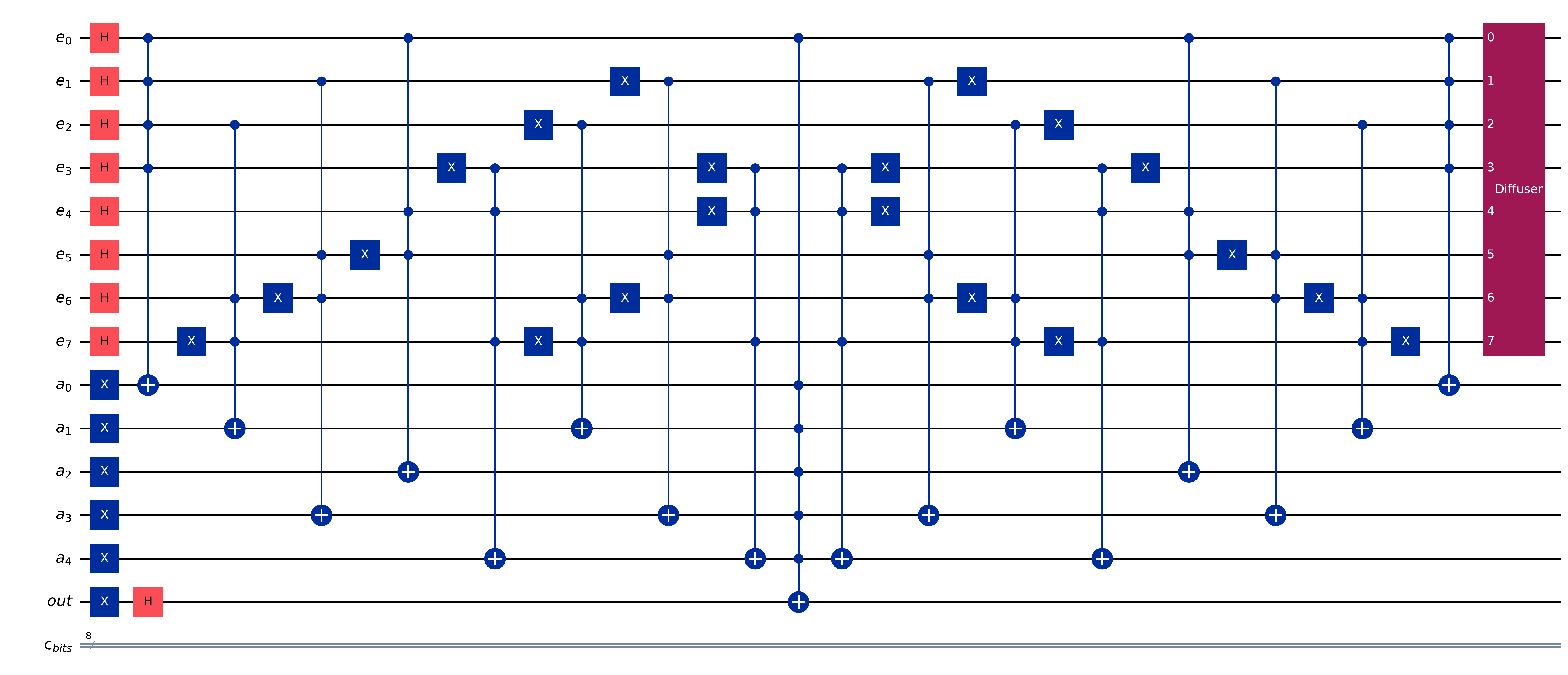}
\caption{Quantum circuit for the four-eloop topology with eight edges (Fig.~\ref{fig:Topologies1}D). The output is 39 causal configurations out of 256 states.  
\label{fig:qc4eloopsA}}
\end{figure}

In the analysis of the four-eloop scenario, Fig.~\ref{fig:Topologies1}D, the eloop clauses specified in Eq.~\ref{eq:clauses4} are implemented into the quantum circuit with eight multicontrolled Toffoli gates: a single multicontrolled Toffoli gate for $a_0^{(4)}$ and $a_2^{(4)}$, and two multicontrolled Toffoli gates for $a_1^{(4)}$, $a_3^{(4)}$, and $a_4^{(4)}$. In addition, the oracle operator requires ten 
NOT gates, as shown in Fig.~\ref{fig:qc4eloopsA}. As for the three-eloop case, these oracle operations are also applied in reversed order. The edge configuration of this topology allows several multicontrolled Toffoli gates to act on distinct qubits. This arrangement enables the simultaneous execution of certain operations within the same time-step, which explains the lower quantum circuit depth.

An important feature of multicontrolled Toffoli gates to be discussed is transpilation. The transpilation process consists of compiling a given quantum circuit to match the specific topology and native gate set of a particular quantum device hardware, as well as its optimisation in order to run on noisy intermediate scale quantum (NISQ) era computers. In the quantum circuit model, high-level quantum circuits are constructed using unitary operators that are implemented as quantum gates. When deploying these circuits on actual quantum hardware, however, they must undergo a decomposition of the unitary quantum gates into the native gate set of the target quantum device. Additionally, optimisation processes are employed to minimise the noise effects and enhance the accuracy of the output results.

\begin{figure}[tb!]
    \centering
    \begin{subfigure}[b]{0.16\textwidth}
        \centering
        \includegraphics[scale=0.38]{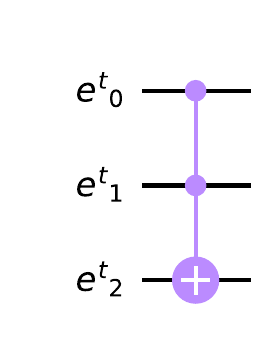}
        \caption{Toffoli gate}
        \label{fig:subfig_a}
    \end{subfigure}
    \hfill
    \begin{subfigure}[b]{0.80\textwidth}
        \centering
        \includegraphics[scale=0.37]{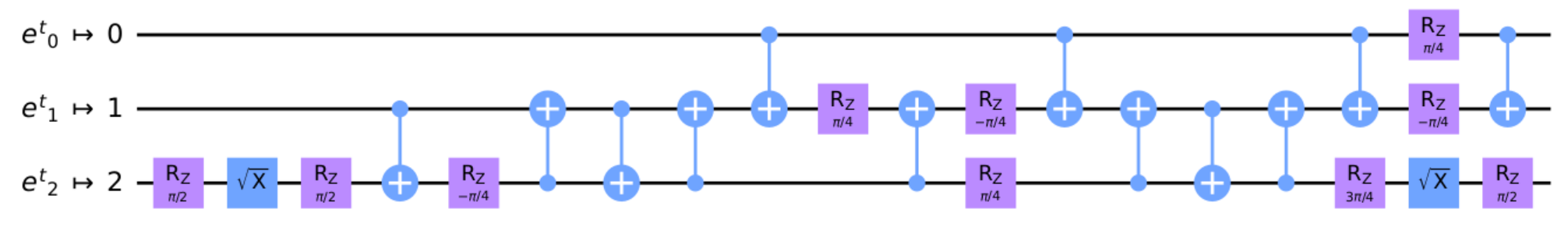}
        \caption{Transpiled Toffoli gate}
        \label{fig:subfig_b}
    \end{subfigure}
    \caption{
      (a) Quantum circuit representation for the Toffoli gate. 
      (b) Equivalent quantum circuit of the Toffoli gate transpiled to the \texttt{ibmq\_belem} device from IBMQ.
    }
    \label{fig:Toffoli_trans}
\end{figure}

The impact of transpilation for a Toffoli gate on the quantum circuit depth is illustrated in Fig~\ref{fig:Toffoli_trans}. The Toffoli gate in Fig~\ref{fig:Toffoli_trans} operates on three qubits: two control qubits and one target qubit. For our analysis, we employed the 5-qubit quantum processor \texttt{ibmq\_belem} (which is now retired\footnote{\href{https://docs.quantum.ibm.com/guides/retired-qpus}{\texttt{https://docs.quantum.ibm.com/guides/retired-qpus}}}; however, a simulated fake backend\footnote{\href{https://docs.quantum.ibm.com/api/qiskit-ibm-runtime/fake_provider}{\texttt{https://docs.quantum.ibm.com/api/qiskit-ibm-runtime/fake\_provider}}} remains accessible). This quantum processor contains a native gate set consisting on $I$, $R_Z$, $SX$, $X$, CNOT and reset gates. The theoretical quantum circuit depth of the Toffoli gate shown in Fig.~\ref{fig:Toffoli_trans}a is one, however its transpilation for the quantum hardware results in a quantum circuit depth of nineteen as illustrated in Fig.~\ref{fig:Toffoli_trans}b. The transpilation of each Toffoli gate increases the quantum circuit depth and therefore has an impact on quantum noise and gate errors.

\begin{table}[tb!]
    \centering
    \footnotesize
    \begin{tabular}{lcccccc} 
    \toprule
    \multicolumn{5}{c}{Four-eloop topology of Fig.~\ref{fig:Topologies1}D}\\ 
    \midrule
    \textbf{Quantum device (qubits)} & \multicolumn{2}{c}{\textbf{Transpiled quantum circuit depth}} & \multicolumn{2}{c}{\textbf{Two-qubit gates}}  \\ 
     & \textbf{BC} & \textbf{MCX} & \textbf{BC} & \textbf{MCX} & \\ 
    \midrule
    \ttfamily FakeMelbourneV2 \rm (14)  & not enough qubits & 1909 & not enough qubits & 1562 & \\ 
    \ttfamily FakeMumbaiV2 \rm (27)     & 1495 & 2163 & 1484 & 1762 & \\ 
    \ttfamily ibm\_brisbane \rm (127)   & 4188 & 5695 & 1558 & 1748 & \\ 
    \ttfamily ibm\_kyoto \rm (127)      & 4726 & 5969 & 1558 & 1748 & \\ 
    \ttfamily ibm\_osaka \rm (127)      & 4672 & 5043 & 1558 & 1748 & \\ 
    \ttfamily ibm\_sherbrooke \rm (127) & 4411 & 5334 & 1558 & 1748 & \\ 
    \bottomrule
    \end{tabular}
    \caption{Key metrics for the transpilation of the quantum circuit corresponding to the four-eloop topology of Fig.~\ref{fig:Topologies1}D in different quantum devices for the binary-clause~(BC) and multicontrolled Toffoli~(MCX) algorithms.
    }
    \label{tb:devices}
\end{table}
To analyse the impact of transpilation, we focus on the four-eloop topology of Fig~\ref{fig:Topologies1}D. We study the transpiled quantum circuit depth for relevant metrics in different quantum backends, both simulated and real quantum hardware, as shown in \Cref{tb:devices}.
The transpiled quantum circuit depth is quite backend-dependent, and is larger for the multicontrolled Toffoli algorithm than for the binary-clause algorithm. The former also requires more two-qubit gates. However, for quantum backends with a limited number of available qubits, only the multicontrolled Toffoli algorithm is capable of simulating complex graph topologies.

It is important to highlight that transpilation may involve more qubits than formally required in the design of the quantum circuit. To address this concern we introduce a new quantum metric, the quantum circuit area, which provides a better understanding of algorithmic efficiency in both quantum simulators and quantum hardware.
The quantum circuit area is defined as the product of the quantum circuit depth and the number of qubits required in the transpilation. This metric provides a more comprehensive evaluation depending in both requirements, the quantum depth of the circuit and its qubit usage, yielding a more complete measure of the computational complexity of the quantum circuit.

\begin{figure}
    \centering
    \includegraphics[width=0.85\linewidth]{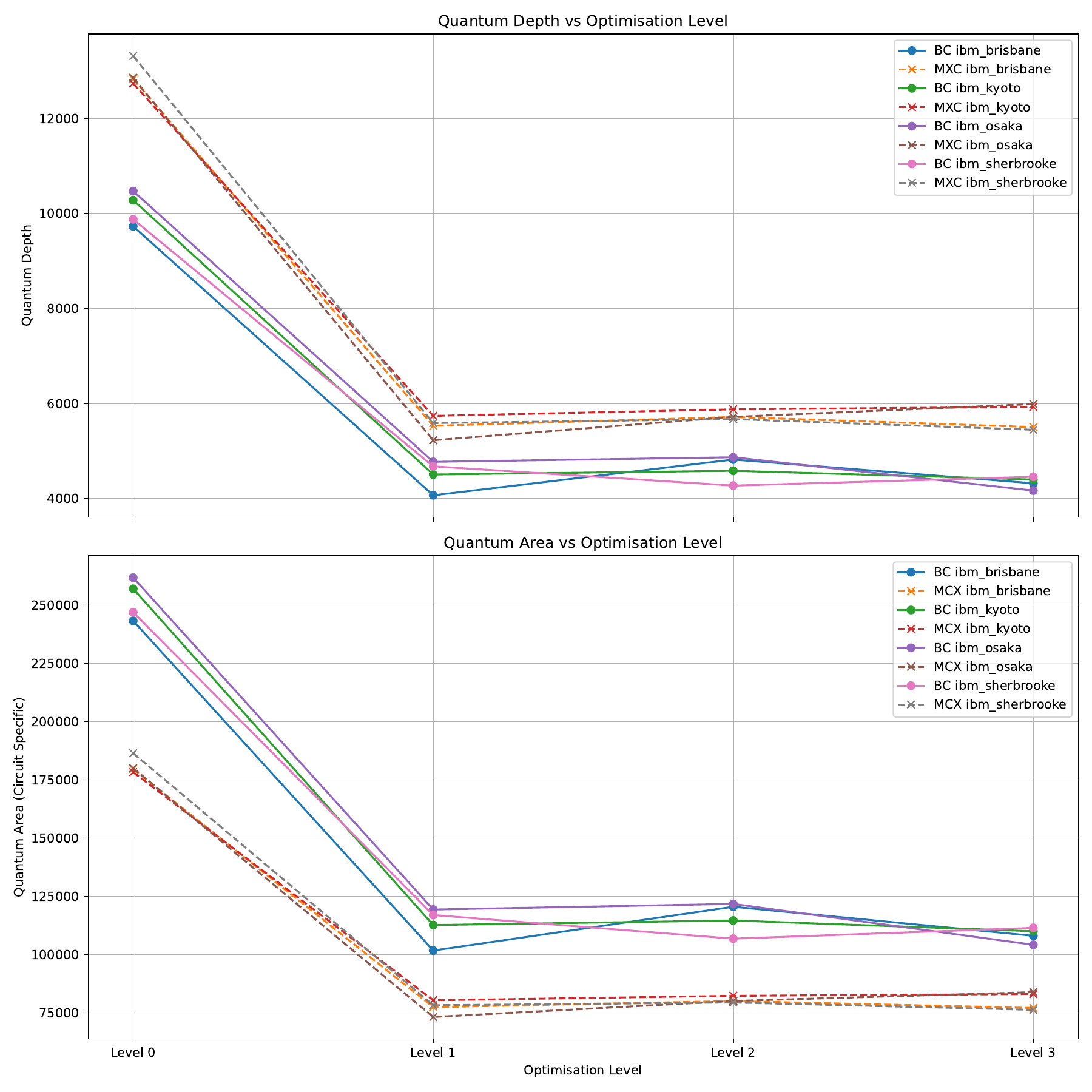}
    \caption{Quantum circuit depth (upper plot) and quantum circuit area (bottom plot) for the four-eloop topology of Fig~\ref{fig:Topologies1}D using the binary-clause (BC) and the multicontrolled Toffoli (MCX) algorithms for various optimisation levels on currently available quantum hardware backends. }
    \label{fig:circuit_specific_quantum_depth-area}
\end{figure}

Continuing the analysis of the four-eloop topology shown in Fig~\ref{fig:Topologies1}D, we present in \Cref{fig:circuit_specific_quantum_depth-area} the transpiled quantum circuit depth and the transpiled quantum circuit area as a function of the optimisation levels on different quantum hardware for both algorithms. The results show a similar evolution in terms of the optimisation levels. Despite the fact that the quantum circuit depth in \Cref{fig:circuit_specific_quantum_depth-area} is higher for the multicontrolled Toffoli algorithm, the quantum circuit area is much smaller compared to  the binary-clause algorithm. 
Therefore, the proposed algorithm manifests a better balance between quantum circuit depth and qubit count, allowing for an improvement in quantum resource efficiency.

\section{Multiloop topologies with multiple number of edges in each edge set}
\label{sec:Application}

\begin{figure}[b]
    \centering
    \includegraphics[scale=0.42]{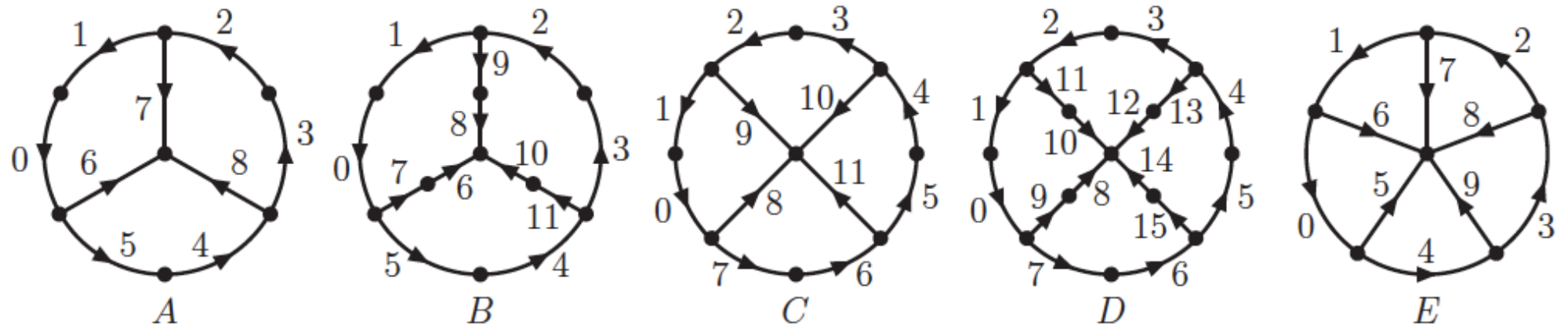}
    \caption{From left to right: three eloops with six and nine vertices, four eloops with a four-point interaction and twelve or sixteen vertices, and five eloops with a five-point interaction. External momenta are attached to the interaction vertices.
    \label{fig:Topologies2}} 
\end{figure}

In the previous section, we have shown a significant reduction in the total number of qubits required to implement the quantum algorithm proposed in this work. 
This achievement allows us to explore more complex topologies than those affordable with the algorithm presented in Ref.~\cite{Ramirez-Uribe:2021ubp}, due to the limitations of the current quantum simulators in terms of the number of qubits. In this section, we inquire into the implementation of the proposed algorithm to involved three-, four- and five-eloop topologies; the number of qubits required for an arbitrary number of edges per edge set is shown in Table~\ref{tb:newtopogral}. The particular cases to explore are illustrated in Fig.~\ref{fig:Topologies2}. The Table~\ref{tb:compa} shows the total number of qubits and quantum circuit depth comparing the algorithm proposed in this work with the one in Ref.~\cite{Ramirez-Uribe:2021ubp}. This table also provides the mixing angle and the probability of the causal states. 
Additionally, Table~\ref{tb:newtopo} compiles the details for the algorithm implementation, number of edges, eloop clauses to test in the oracle operator, and Toffoli and NOT gates required for the oracle operator. It is important to remark that from the multiloop topologies depicted in Fig.~\ref{fig:Topologies2}, only the three-eloop topology with nine edges and the simplest five-eloop are formally reachable with the algorithm in Ref.~\cite{Ramirez-Uribe:2021ubp} in the Qiskit quantum simulator.

\begin{table}
    \begin{center}
    \begin{tabular}{lccl} \toprule  
    \bfseries eloops & $\ket{e}$ & $\ket{a}$ & Total Qubits\\ \hline
    three & $n$ or $n+1$  & $4$ to $7\ph{0}$ & $n+(5$ to $9)$\\
    four$^{(c,t,s,u)}$ & $n$ or $n+1$ &  $5$ to $13$ & $n + (6$ to $15)$\\
    five$^{(c)}$ & $n$ or $n+1$ & $6$ to $21$ & $n+(6\text{ to } 22)$ \\
    \bottomrule 
    \end{tabular}
    \end{center}
    \caption{Number of qubits required for an arbitrary three-, four- and five-eloop configuration in Fig.~\ref{fig:Topologies2}. The total number of qubits includes the oracle's marker. 
    \label{tb:newtopogral}}
\end{table}

\begin{table}[t] \small 
    \begin{center} 
    \begin{tabular}{clccccr} 
    \toprule 
    \bfseries 
    Fig. & \bfseries eloops(edges) & $\substack{\text{Total} \\ \text{Qubits}}$ & $\substack{\text{Quantum} \\ \text{Depth}}$ & $\theta$ &  $\sin^2(\theta_t)$ \\ \midrule 
    \ref{fig:Topologies2}A & three ($9$) & $14~|~25$ & $18~|~27$ & $35.2^\circ$  & $0.93$ \\
    \ref{fig:Topologies2}B & three($12$) & $21~|~36$ & $32~|~33$ & $28.0^\circ$ & $0.99$ \\ 
    \ref{fig:Topologies2}C & four$^{(c)}$($12$) & $18~|~33$ & $16~|~31$ & $32.8^\circ$  & $0.98$  \\  
    \ref{fig:Topologies2}D & four$^{(c)}$($16$) & $31~|~50$ & $46~|~53$ & $27.7^\circ$ & $0.98$  \\  
    \ref{fig:Topologies2}E & five$^{(c)}$($10$) & $17~|~31$ &$24~|~25$ & $28.9^\circ$ & $0.99$  \\
    \bottomrule 
    \end{tabular}
    \end{center}
    \caption{Quantum resources required and quantum circuit depth for the quantum algorithm implementation of the multiloop topologies in Fig.~\ref{fig:Topologies2}. The first and second numbers in the third and fourth columns correspond to the algorithm proposed in this work and to the algorithm in Ref.~\cite{Ramirez-Uribe:2021ubp}, respectively. The mixing angle and the probability of the causal states after amplitude amplification are the same for both algorithms.}
    \label{tb:compa} 
\end{table}

\begin{table}[t] \small 
    \begin{center} 
    \begin{tabular}{clccccccrr} 
    \toprule 
    \bfseries 
    Fig. & \bfseries eloops(edges) & $\ket{e}$ & $\ket{a}$ &  $\substack{\text{Toffoli} \\ \text{Gates}}$ & $\substack{\text{NOT} \\ \text{Gates}}$ &
    $\substack{\text{Causal} \\ \text{States}}$ &
    $\substack{\text{Total} \\ \text{states}}$\\ \midrule 
    \ref{fig:Topologies2}A & three ($9$) & $9$ & $4$ & $14$ & $45$ & $170$ & $512$ \\
    \ref{fig:Topologies2}B & three($12$) & $13$ & $7$ & $21$ & $48$ & $1804$ & $8192$ \\ 
    \ref{fig:Topologies2}C & four$^{(c)}$($12$) & $12$ & $5$ & $17$ & $66$ & $1199$ & $4096$ \\  
    \ref{fig:Topologies2}D & four$^{(c)}$($16$) & $17$ & $13$ & $39$ & $85$ & $28343$ & $131072$\\  
    \ref{fig:Topologies2}E & five$^{(c)}$($10$) & $10$ & $6$ & $25$  & $37$ & $240$ & $1024$\\
    \bottomrule 
    \end{tabular}
    \end{center}
    \caption{Number of qubits required in the $\ket{e}$ and $\ket{a}$ registers for the algorithm implementation of the multiloop topologies depicted in Fig.~\ref{fig:Topologies2}, number of Toffoli and $X$ gates, number of causal states and total number of states.}
    \label{tb:newtopo} 
\end{table}

\subsection{Three eloops with nine and twelve edges}
\label{ssec:3eloops}
    
\begin{figure}[tb!]
\centering
\includegraphics[width=\textwidth]{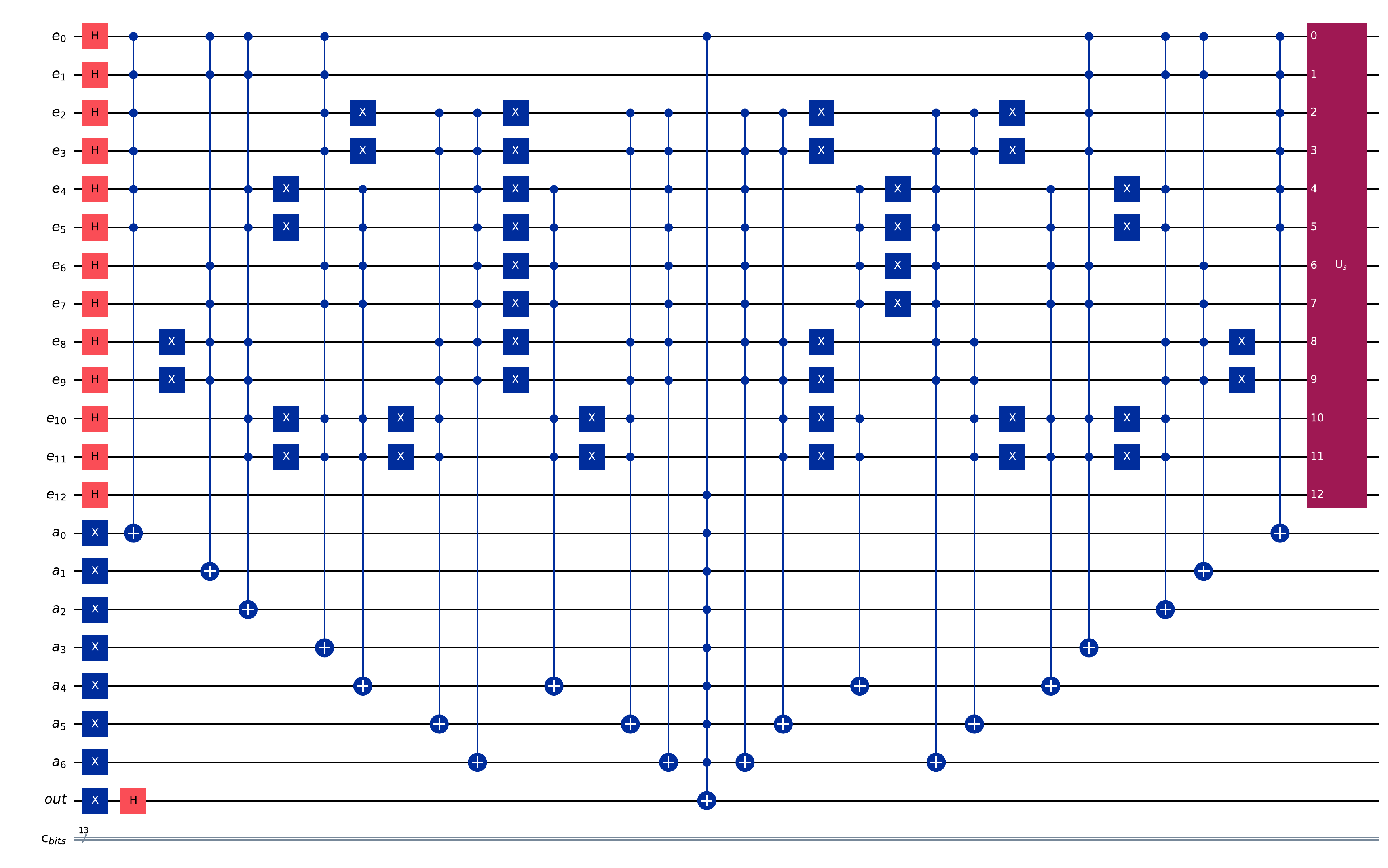}
\caption{Quantum circuit for the three-eloop topology with twelve edges (Fig.~\ref{fig:Topologies2}B). The output is 1804 causal configurations out of 8192 states. The last qubit in the $\ket{e}$ registed is an ancillary qubit used to adjust the initial mixing angle to the optimal value.
\label{fig:qc3eloops}}
\end{figure}
    
At three eloops, we work with the selected multiloop topologies shown in Figs.~\ref{fig:Topologies2}A and~\ref{fig:Topologies2}B. The first three-eloop topolology (Fig.~\ref{fig:Topologies2}A) considers two edges in the external sets of edges and one edge in the internal sets of edges, and the Boolean functions are given by,
\beq \label{eq:3eloop2edges-a}
    s_i=2i\wedge (2i+1),\quad i\in \{0,1,2\} \quad \text{and} \quad s_i=3+i, \quad i\in \{3,4,5\}~.
\eeq
Regarding the eloop clauses, it requires the four eloop clauses in \Eq{eq:clauses3}.

For the second three-eloop topology (Fig.~\ref{fig:Topologies2}B), we take two edges in all the sets of edges, and the Boolean functions are given by,
\beq \label{eq:3eloop2edges-b}
    s_i=2i\wedge (2i+1), \qquad i\in \{0, \ldots,5 \}~.
\eeq
It requires, in addition to the eloop clauses in \Eq{eq:clauses3}, the following eloop clauses to test acyclicity over four sets of edges
\begin{align}
    a_4^{(3)} &= \toff (s_0 \land s_1 \land s_3 \land \bar{s}_5) ~, &
    a_5^{(3)} &= \Toff (s_1 \land s_2 \land \bar{s}_3 \land s_4) ~, \nn\\
    a_6^{(3)} &= \toff (s_0 \land s_2 \land \bar{s}_4 \land s_5) ~. &
\end{align}

The total number of qubits required to implement the configurations in \Eq{eq:3eloop2edges-a} and \Eq{eq:3eloop2edges-b} is fourteen and twenty-one respectively, and the quantum circuit depth results in eighteen and thirty-two, respectively. In the binary-clause approach these configurations require twenty-five and thirty-six qubits, and the quantum circuit depths would be twenty-seven and thirty-three, respectively; the latter exceeding the number of qubits available on the Qiskit simulator. The quantum circuit implemented for the configuration in~\Eq{eq:3eloop2edges-b} is shown in Fig.~\ref{fig:qc3eloops}.

Regarding the number of eloop clauses, any three-eloop topology with multiple edges located exclusively in the external sets of edges do not enlarge the minimum number of clauses to be tested, whereas including two or more edges in every edge set requires to test the maximum number of eloops clauses. In fact, thirty-two is the maximum quantum circuit depth that a three-eloop topology can have due to the fact that the quantum circuit depth of the proposed quantum algorithm increases with the number of eloop clauses but it does not scale with additional edges.

\subsection{Four eloops with twelve and sixteen edges}
\label{ssec:4eloops}

\begin{figure}[tb!]
\includegraphics[width=\textwidth]{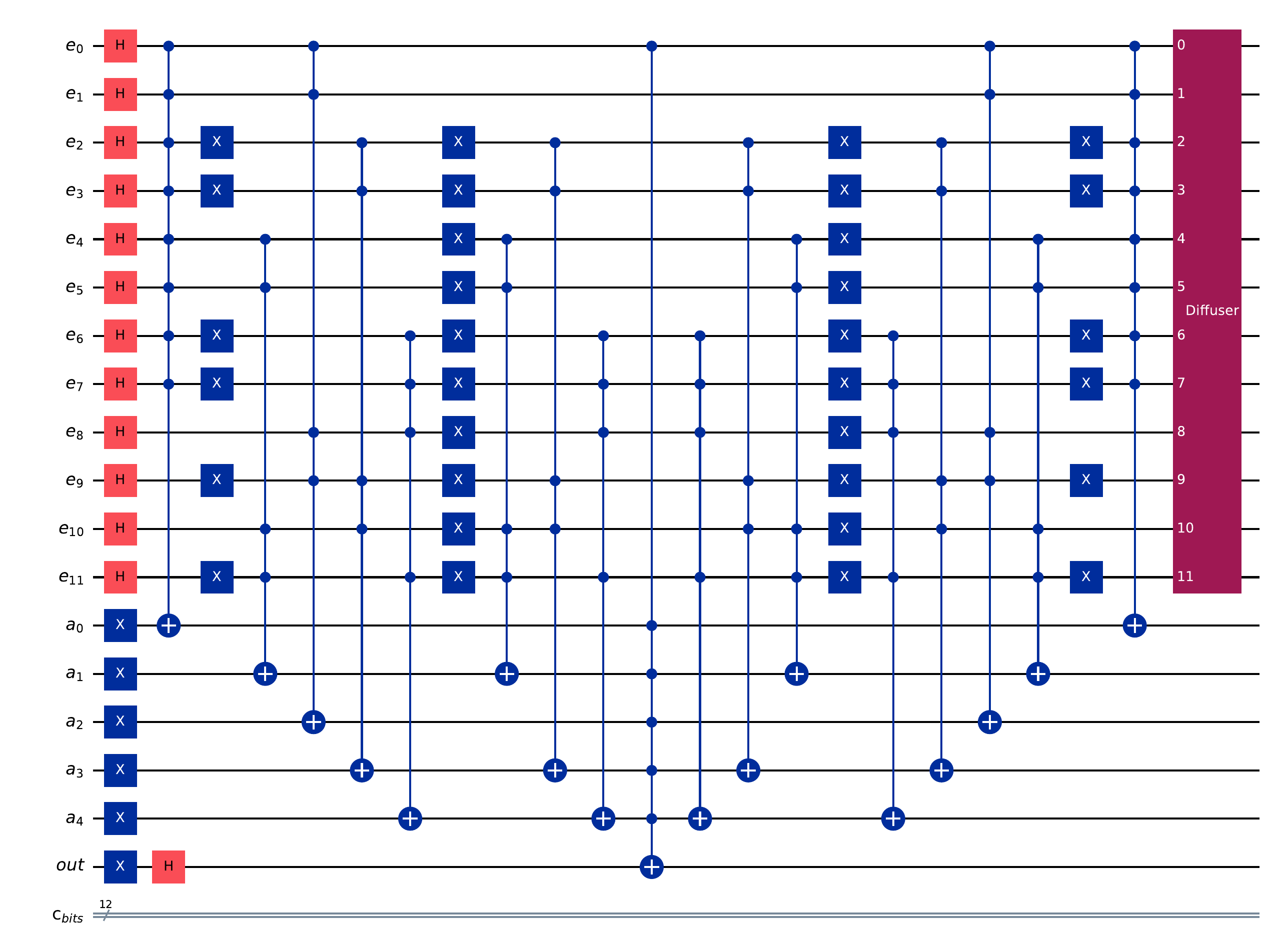}
\caption{Quantum circuit for the four-eloop topology in Fig.~\ref{fig:Topologies2}C. The output is 1199 causal configurations out of 4096 states. 
\label{fig:qc4eloops}}

\end{figure}

The four-eloop examples to analyse are those depicted in Figs.~\ref{fig:Topologies2}C and~\ref{fig:Topologies2}D. The first loop topology (Fig.~\ref{fig:Topologies2}C) considers two edges in the external sets of edges and one edge in the internal sets of edges, and are given by
\beq \label{eq:4eloop2edges-a}
    s_i=2i\wedge (2i+1),\quad i\in \{0,1,2,3\} \quad \text{and} \quad s_i=3+i, \quad i\in \{4,5,6,7\}~.
\eeq
The second loop topology (Fig.~\ref{fig:Topologies2}D), consists in two edges in every edge set,
\beq \label{eq:4eloop2edges-b}
    s_i=2i\wedge (2i+1), \qquad i\in \{0, \ldots,7 \}~.
\eeq
The complete set of eloop clauses for the four-eloop topology of Fig.~\ref{fig:Topologies2}D, together with the eloop clauses from Eq.~\eqref{eq:clauses4} are given by 
\begin{align}
    a_5^{(4)} &= \toff (s_0 \land s_1 \land s_4 \land \bar{s}_6) ~, &
    a_6^{(4)} &= \Toff (s_1 \land s_2 \land s_5 \land \bar{s}_7) ~, \nn\\
    a_7^{(4)} &= \Toff (s_2 \land s_3 \land \bar{s}_4 \land s_6) ~, &
    a_8^{(4)} &= \toff (s_0 \land s_3 \land \bar{s}_5 \land s_7) , \nn\\
    a_9^{(4)} &= \toff (s_0 \land s_1 \land s_2 \land s_4 \land \bar{s}_7) ~, &
    a_{10}^{(4)} &= \Toff (s_1 \land s_2 \land s_3 \land \bar{s}_4 \land s_5) ~, \nn\\
    a_{11}^{(4)} &= \toff (s_0 \land s_2 \land s_3 \land \bar{s}_5 \land s_6) ~, &
    a_{12}^{(4)} &= \toff (s_0 \land s_1 \land s_3 \land \bar{s}_6 \land s_7) ~. 
\label{eq:4eloopextra}
\end{align}
The implementation of the algorithm for these two cases requires a total number of eighteen and thirty thirty-one qubits, respectively, compared to thirty-three and fifty for the binary-clause approach. In terms of the quantum circuit depth, it is sixteen for Fig.~\ref{fig:Topologies2}C and forty-six for Fig.~\ref{fig:Topologies2}D, hence a maximum quantum circuit depth of forty-six for any four-eloop topology independently of the number of edges. Similar to the previous cases of three eloops, the four-eloop topology associated to Eq.~\eqref{eq:4eloop2edges-a} does not need to test additional eloop clauses besides the minimum requirement. The quantum circuit for Fig.~\ref{fig:Topologies2}C is shown in Fig.~\ref{fig:qc4eloops}.

For the four-eloop topology configuration in Fig.~\ref{fig:Topologies2}D with two edges in all the sets of edges, it is necessary to check additional eloop clauses, considering those associated with sub-eloops generated with four and five edge sets, \Eq{eq:4eloopextra}, giving a total of thirteen eloop clauses for all the causal configurations.

\subsection{Five eloops with ten egdes}

\begin{figure}[tb!]
    \includegraphics[width=\textwidth]{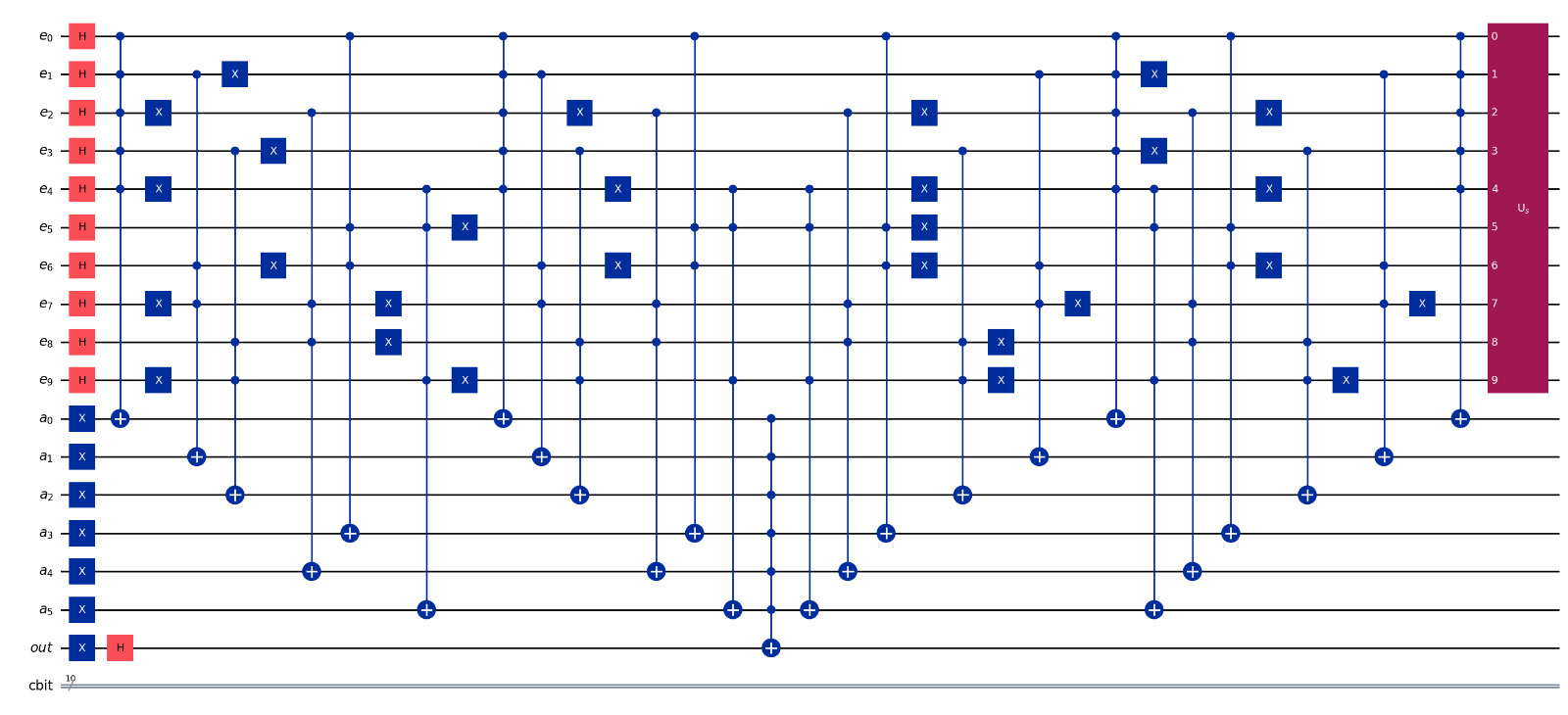}
    \caption{Quantum circuit for the five-eloop topology with one edge per line shown in \Cref{fig:Topologies2}E. The output is 240 causal configurations out of 1024 states.}
    \label{fig:qc5eloops}
\end{figure}
\begin{table}[tb!]
    \centering
    \footnotesize
    \begin{tabular}{lcccccc} 
    \toprule
    \multicolumn{5}{c}{Five-eloop topology, \Cref{fig:Topologies2}E}\\ 
    \midrule
    \textbf{Quantum device (qubits)} & \multicolumn{2}{c}{\textbf{Transpiled quantum depth}} & \multicolumn{2}{c}{\textbf{Two-qubit gates}}  \\ 
     & BC & MCX & BC & MCX & \\ 
    \midrule
    \ttfamily ibm\_brisbane \rm (127)   & 10745 & 13008 & 3629 & 4377 & \\ 
    \ttfamily ibm\_kyiv \rm (127)       & 12410 & 13909 & 3629 & 4377 & \\ 
    \ttfamily ibm\_sherbrooke \rm (127) & 12485 & 13281 & 3629 & 4377 & \\ 
    \bottomrule
    \end{tabular}
    \caption{Key metrics for the transpilation of the quantum circuit corresponding to the four-eloop topology of \Cref{fig:Topologies2}E in different quantum devices for the binary-clause~(BC) and multicontrolled Toffoli~(MCX) algorithms.
    }
    \label{tb:5_BC_MCX}
\end{table}

The last eloop topology we implement has five eloops with a single edge per set $s_i$ which is depicted in Fig.~\ref{fig:Topologies2}E. The eloop clauses to validate are composed by
\begin{align} \label{eq:5_eloop_mcx_clauses}
    a_0^{(5)} &= \Toff (s_0\wedge s_1\wedge s_2\wedge s_3\wedge s_4) ~,\quad&    
    a_1^{(5)} &= \Toff (s_0\wedge s_5\wedge \bar{s}_6) ~, \nn\\
    a_2^{(5)} &= \Toff (s_1\wedge s_6\wedge \bar{s}_7) ~,      \quad&            
    a_3^{(5)} &= \Toff (s_2\wedge s_7\wedge \bar{s}_8) ~, \nn\\
    a_4^{(5)} &= \Toff (s_3\wedge s_8 \wedge \bar{s}_9) ~,     \quad&            
    a_5^{(5)} &= \Toff (s_4\wedge \bar{s}_5\wedge s_9) ~.
\end{align}
The mixing angle is $\theta = 28.9^{\circ}$ without the need of fixing a qubit from the register $\ket{e}$. This means that the querying probability for finding the 240 causal states out of the $2^{10}$ total states is close to one, $\sin^2(\theta_t) = 0.99$, with a single iteration, $t = 1$. 

In terms of quantum resources, the algorithm based on multi-controlled Toffoli gates requires seventeen qubits to build the corresponding quantum circuit (\Cref{fig:qc5eloops}) with a quantum circuit depth of twenty-three versus thirty-one qubits and a quantum circuit depth of twenty-five. Note that in this case the inclusion of additional edges in the five-eloop topology would prevent the binary-clause algorithm from being implemented in the Qiskit quantum simulator, given its limited capacity of 32 qubits.

Continuing the analysis of the five-eloop topology we follow with the study of the transpiled quantum circuit depth for relevant metrics in different quantum backends, both simulated and real quantum hardware, as shown in \Cref{tb:5_BC_MCX}. 
The transpiled quantum circuit depth is larger and requires more two-qubit gates for the multicontrolled Toffoli algorithm than for the binary-clause algorithm.

\begin{figure}
    \centering
    \includegraphics[width=\linewidth]{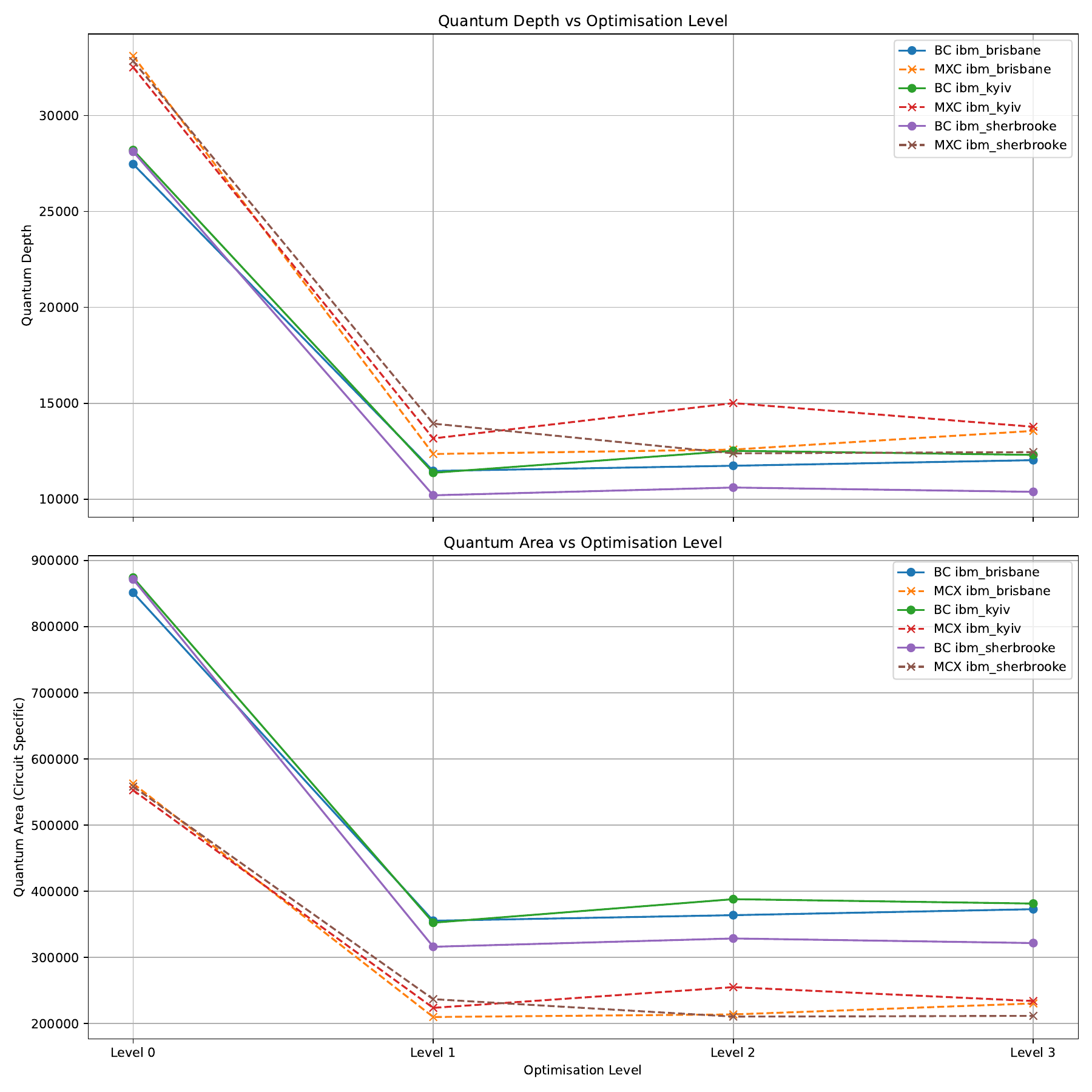}
    \caption{Quantum circuit depth (upper plot) and quantum circuit area (bottom plot) for the five-eloop topology of \Cref{fig:Topologies2}E in terms of optimisation levels for the transpilation. 
    The solid lines represent the transpilation of the BC quantum algorithm and the dashed lines represent the transpilation for the MCX. }
    \label{fig:5eloops_scaling}
\end{figure}

As mentioned in the previous section, to provide a more appropriate assessment of the complexity of the algorithm we use the quantum circuit area considering both, the transpiled quantum depth and qubit usage. In \Cref{fig:5eloops_scaling} we present the transpiled quantum circuit depth and the transpiled quantum circuit area as a function of the optimisation levels on different quantum hardware, noting comparable results with the four-eloop topology. It shows a similar evolution in terms of the optimisation levels, and despite having a transpile quantum depth higher for the MCX algorithm, the quantum circuit area is smaller compared to the BC algorithm.

In addition, the simulations revealed a significant difference in execution time and resource consumption. The time taken in the ideal simulation in Qiskit (noiseless) was around 165 seconds for the BC algorithm and less than 0.1 seconds for the MCX algorithm. The BC algorithm simulation fully utilises the computer processor (100\% CPU load on an Intel i7 processor, running Qiskit local on a Linux operating system) and consumed $\sim90\%$ of the available 64 GB RAM, pushing the limits of the quantum simulator. In contrast, the MCX algorithm simulation was executed with significantly fewer resources. 

\section{Conclusions}
\label{sec:Conclusion}

We have presented an efficient and systematic quantum algorithm for identifying the causal configurations for multiloop Feynman diagrams, which is equivalent to querying DAG configurations of undirected graphs in graph theory. The approach presented in this paper defines an oracle operator using only multicontrolled Toffoli gates and NOT (Pauli-$X$) gates simplifying the design and encoding of the quantum circuits. Furthermore, it allows us to significantly reduce the total number of qubits required and the theoretical quantum circuit depth with respect to a binary-clause algorithm.

We have also analysed the quantum complexity of the transpiled quantum circuits. In this case, the quantum circuit area is the most appropriate metric to assess the complexity of the quantum algorithm due to the fact that jointly evaluates the number of qubits and the transpiled quantum depth. 
The quantum circuit area of the proposed algorithm is smaller than that of the binary-clause algorithm, despite having a greater transpiled quantum circuit depth, thus showing a relevant improvement in quantum resource efficiency. The reduction in the number of qubits required has allowed us to explore complex multiloop topologies using the IBM Quantum platform; in particular, we have explicitly implemented three eloops with six and nine vertices, four eloops with twelve and sixteen vertices, and five eloops with a five point interaction and ten edges, successfully identifying their causal configurations.

\section*{Acknowledgements}
We are very grateful to Fundaci\'on Centro Tecnol\'ogico de la Informaci\'on y la Comunicaci\'on (CTIC) for granting us access to their simulator \emph{Quantum Testbed} (QUTE). We thank also access to IBMQ. This work is supported by the Spanish Government (Agencia Estatal de Investigaci\'on MCIN/AEI/ 10.13039/501100011033) Grant No. PID2020-114473GB-I00, No. PID2023-146220NB-I00 and No. CEX2023-001292-S, and Generalitat Valenciana Grant No. PROMETEO/2021/071 and ASFAE/2022/009 (Planes Complementarios de I+D+i, NextGenerationEU). This work is also supported by the  Ministry of Economic Affairs and Digital Transformation of the Spanish Government and NextGenerationEU through the Quantum Spain project, and by CSIC Interdisciplinary Thematic Platform (PTI+) on Quantum Technologies (PTI-QTEP+). 
SRU acknowledges support from CONAHCyT through Project No.~320856 (Paradigmas y Controversias de la Ciencia 2022), Ciencia de Frontera 2021-2042 and Sistema Nacional de Investigadores; AERO from the Spanish Government (PRE2018-085925).

\bibliographystyle{JHEP}
\bibliography{main}

\providecommand{\href}[2]{#2}\begingroup\raggedright\begin{thebibliography}{10}

\bibitem{Feynman:1981tf}
R.~P. Feynman, \emph{{Simulating physics with computers}},
  \href{http://dx.doi.org/10.1007/BF02650179}{\emph{Int. J. Theor. Phys.} {\bf
  21} (1982) 467--488}.

\bibitem{Humble:2022klb}
T.~S. Humble, G.~N. Perdue and M.~J. Savage, \emph{{Snowmass Computational
  Frontier: Topical Group Report on Quantum Computing}},
  \href{http://arxiv.org/abs/2209.06786}{{\tt 2209.06786}}.

\bibitem{Rodrigo:2024say}
G.~Rodrigo, \emph{{Quantum Algorithms in Particle Physics}},
  \href{http://dx.doi.org/10.5506/APhysPolBSupp.17.2-A14}{\emph{Acta Phys.
  Polon. Supp.} {\bf 17} (2024) 2--A14},
  [\href{http://arxiv.org/abs/2401.16208}{{\tt 2401.16208}}].

\bibitem{Jordan:2011ne}
S.~P. Jordan, K.~S.~M. Lee and J.~Preskill, \emph{{Quantum Algorithms for
  Quantum Field Theories}},
  \href{http://dx.doi.org/10.1126/science.1217069}{\emph{Science} {\bf 336}
  (2012) 1130--1133}, [\href{http://arxiv.org/abs/1111.3633}{{\tt 1111.3633}}].

\bibitem{Banuls:2019bmf}
M.~C. Ba\~nuls et~al., \emph{{Simulating Lattice Gauge Theories within Quantum
  Technologies}},
  \href{http://dx.doi.org/10.1140/epjd/e2020-100571-8}{\emph{Eur. Phys. J. D}
  {\bf 74} (2020) 165}, [\href{http://arxiv.org/abs/1911.00003}{{\tt
  1911.00003}}].

\bibitem{Zohar:2015hwa}
E.~Zohar, J.~I. Cirac and B.~Reznik, \emph{{Quantum Simulations of Lattice
  Gauge Theories using Ultracold Atoms in Optical Lattices}},
  \href{http://dx.doi.org/10.1088/0034-4885/79/1/014401}{\emph{Rept. Prog.
  Phys.} {\bf 79} (2016) 014401}, [\href{http://arxiv.org/abs/1503.02312}{{\tt
  1503.02312}}].

\bibitem{Byrnes:2005qx}
T.~Byrnes and Y.~Yamamoto, \emph{{Simulating lattice gauge theories on a
  quantum computer}},
  \href{http://dx.doi.org/10.1103/PhysRevA.73.022328}{\emph{Phys. Rev. A} {\bf
  73} (2006) 022328}, [\href{http://arxiv.org/abs/quant-ph/0510027}{{\tt
  quant-ph/0510027}}].

\bibitem{Magano:2021jzd}
D.~Magano et~al., \emph{{Quantum speedup for track reconstruction in particle
  accelerators}},
  \href{http://dx.doi.org/10.1103/PhysRevD.105.076012}{\emph{Phys. Rev. D} {\bf
  105} (2022) 076012}, [\href{http://arxiv.org/abs/2104.11583}{{\tt
  2104.11583}}].

\bibitem{Duckett:2022ccc}
P.~Duckett, G.~Facini, M.~Jastrzebski, S.~Malik, T.~Scanlon and S.~Rettie,
  \emph{{Reconstructing charged particle track segments with a quantum-enhanced
  support vector machine}},
  \href{http://dx.doi.org/10.1103/PhysRevD.109.052002}{\emph{Phys. Rev. D} {\bf
  109} (2024) 052002}, [\href{http://arxiv.org/abs/2212.07279}{{\tt
  2212.07279}}].

\bibitem{Schwagerl:2023elf}
T.~Schw\"agerl, C.~Issever, K.~Jansen, T.~J. Khoo, S.~K\"uhn, C.~T\"uys\"uz
  et~al., \emph{{Particle track reconstruction with noisy intermediate-scale
  quantum computers}},  \href{http://arxiv.org/abs/2303.13249}{{\tt
  2303.13249}}.

\bibitem{Wei:2019rqy}
A.~Y. Wei, P.~Naik, A.~W. Harrow and J.~Thaler, \emph{{Quantum Algorithms for
  Jet Clustering}},
  \href{http://dx.doi.org/10.1103/PhysRevD.101.094015}{\emph{Phys. Rev. D} {\bf
  101} (2020) 094015}, [\href{http://arxiv.org/abs/1908.08949}{{\tt
  1908.08949}}].

\bibitem{Pires:2020urc}
D.~Pires, Y.~Omar and J.~a. Seixas, \emph{{Adiabatic quantum algorithm for
  multijet clustering in high energy physics}},
  \href{http://dx.doi.org/10.1016/j.physletb.2023.138000}{\emph{Phys. Lett. B}
  {\bf 843} (2023) 138000}, [\href{http://arxiv.org/abs/2012.14514}{{\tt
  2012.14514}}].

\bibitem{deLejarza:2022bwc}
J.~J.~M. de~Lejarza, L.~Cieri and G.~Rodrigo, \emph{{Quantum clustering and jet
  reconstruction at the LHC}},
  \href{http://dx.doi.org/10.1103/PhysRevD.106.036021}{\emph{Phys. Rev. D} {\bf
  106} (2022) 036021}, [\href{http://arxiv.org/abs/2204.06496}{{\tt
  2204.06496}}].

\bibitem{deLejarza:2022vhe}
J.~J.~M. de~Lejarza, L.~Cieri and G.~Rodrigo, \emph{{Quantum jet clustering
  with LHC simulated data}},
  \href{http://dx.doi.org/10.22323/1.414.0241}{\emph{PoS} {\bf ICHEP2022}
  (2022) 241}, [\href{http://arxiv.org/abs/2209.08914}{{\tt 2209.08914}}].

\bibitem{Barata:2021yri}
J.~a. Barata and C.~A. Salgado, \emph{{A quantum strategy to compute the jet
  quenching parameter $\hat{q}$}},
  \href{http://dx.doi.org/10.1140/epjc/s10052-021-09674-9}{\emph{Eur. Phys. J.
  C} {\bf 81} (2021) 862}, [\href{http://arxiv.org/abs/2104.04661}{{\tt
  2104.04661}}].

\bibitem{Barata:2022wim}
J.~a. Barata, X.~Du, M.~Li, W.~Qian and C.~A. Salgado, \emph{{Medium induced
  jet broadening in a quantum computer}},
  \href{http://dx.doi.org/10.1103/PhysRevD.106.074013}{\emph{Phys. Rev. D} {\bf
  106} (2022) 074013}, [\href{http://arxiv.org/abs/2208.06750}{{\tt
  2208.06750}}].

\bibitem{Barata:2023clv}
J.~a. Barata, X.~Du, M.~Li, W.~Qian and C.~A. Salgado, \emph{{Quantum
  simulation of in-medium QCD jets: Momentum broadening, gluon production, and
  entropy growth}},
  \href{http://dx.doi.org/10.1103/PhysRevD.108.056023}{\emph{Phys. Rev. D} {\bf
  108} (2023) 056023}, [\href{http://arxiv.org/abs/2307.01792}{{\tt
  2307.01792}}].

\bibitem{Bauer:2019qxa}
C.~W. Bauer, W.~A. de~Jong, B.~Nachman and D.~Provasoli, \emph{{Quantum
  Algorithm for High Energy Physics Simulations}},
  \href{http://dx.doi.org/10.1103/PhysRevLett.126.062001}{\emph{Phys. Rev.
  Lett.} {\bf 126} (2021) 062001}, [\href{http://arxiv.org/abs/1904.03196}{{\tt
  1904.03196}}].

\bibitem{Bauer:2021gup}
C.~W. Bauer, M.~Freytsis and B.~Nachman, \emph{{Simulating Collider Physics on
  Quantum Computers Using Effective Field Theories}},
  \href{http://dx.doi.org/10.1103/PhysRevLett.127.212001}{\emph{Phys. Rev.
  Lett.} {\bf 127} (2021) 212001}, [\href{http://arxiv.org/abs/2102.05044}{{\tt
  2102.05044}}].

\bibitem{Bepari:2020xqi}
K.~Bepari, S.~Malik, M.~Spannowsky and S.~Williams, \emph{{Towards a quantum
  computing algorithm for helicity amplitudes and parton showers}},
  \href{http://dx.doi.org/10.1103/PhysRevD.103.076020}{\emph{Phys. Rev. D} {\bf
  103} (2021) 076020}, [\href{http://arxiv.org/abs/2010.00046}{{\tt
  2010.00046}}].

\bibitem{Williams:2021lvr}
K.~Bepari, S.~Malik, M.~Spannowsky and S.~Williams, \emph{{Quantum walk
  approach to simulating parton showers}},
  \href{http://dx.doi.org/10.1103/PhysRevD.106.056002}{\emph{Phys. Rev. D} {\bf
  106} (2022) 056002}, [\href{http://arxiv.org/abs/2109.13975}{{\tt
  2109.13975}}].

\bibitem{Perez-Salinas:2020nem}
A.~P\'erez-Salinas, J.~Cruz-Martinez, A.~A. Alhajri and S.~Carrazza,
  \emph{{Determining the proton content with a quantum computer}},
  \href{http://dx.doi.org/10.1103/PhysRevD.103.034027}{\emph{Phys. Rev. D} {\bf
  103} (2021) 034027}, [\href{http://arxiv.org/abs/2011.13934}{{\tt
  2011.13934}}].

\bibitem{Cruz-Martinez:2023vgs}
J.~M. Cruz-Martinez, M.~Robbiati and S.~Carrazza, \emph{{Multi-variable
  integration with a variational quantum circuit}},
  \href{http://dx.doi.org/10.1088/2058-9565/ad5866}{\emph{Quantum Sci.
  Technol.} {\bf 9} (2024) 035053},
  [\href{http://arxiv.org/abs/2308.05657}{{\tt 2308.05657}}].

\bibitem{Chawdhry:2023jks}
H.~A. Chawdhry and M.~Pellen, \emph{{Quantum simulation of colour in
  perturbative quantum chromodynamics}},
  \href{http://dx.doi.org/10.21468/SciPostPhys.15.5.205}{\emph{SciPost Phys.}
  {\bf 15} (2023) 205}, [\href{http://arxiv.org/abs/2303.04818}{{\tt
  2303.04818}}].

\bibitem{deJong:2020tvx}
W.~A. De~Jong, M.~Metcalf, J.~Mulligan, M.~P\l{}osko\'n, F.~Ringer and X.~Yao,
  \emph{{Quantum simulation of open quantum systems in heavy-ion collisions}},
  \href{http://dx.doi.org/10.1103/PhysRevD.104.L051501}{\emph{Phys. Rev. D}
  {\bf 104} (2021) 051501}, [\href{http://arxiv.org/abs/2010.03571}{{\tt
  2010.03571}}].

\bibitem{Guan:2020bdl}
W.~Guan, G.~Perdue, A.~Pesah, M.~Schuld, K.~Terashi, S.~Vallecorsa et~al.,
  \emph{{Quantum Machine Learning in High Energy Physics}},
  \href{http://dx.doi.org/10.1088/2632-2153/abc17d}{\emph{Mach. Learn. Sci.
  Tech.} {\bf 2} (2021) 011003}, [\href{http://arxiv.org/abs/2005.08582}{{\tt
  2005.08582}}].

\bibitem{Wu:2020cye}
S.~L. Wu et~al., \emph{{Application of quantum machine learning using the
  quantum variational classifier method to high energy physics analysis at the
  LHC on IBM quantum computer simulator and hardware with 10 qubits}},
  \href{http://dx.doi.org/10.1088/1361-6471/ac1391}{\emph{J. Phys. G} {\bf 48}
  (2021) 125003}, [\href{http://arxiv.org/abs/2012.11560}{{\tt 2012.11560}}].

\bibitem{Trenti:2020ceh}
T.~Felser, M.~Trenti, L.~Sestini, A.~Gianelle, D.~Zuliani, D.~Lucchesi et~al.,
  \emph{{Quantum-inspired machine learning on high-energy physics data}},
  \href{http://dx.doi.org/10.1038/s41534-021-00443-w}{\emph{npj Quantum Inf.}
  {\bf 7} (2021) 111}, [\href{http://arxiv.org/abs/2004.13747}{{\tt
  2004.13747}}].

\bibitem{Herbert:2021xgs}
S.~Herbert, \emph{{Quantum Monte Carlo Integration: The Full Advantage in
  Minimal Circuit Depth}},
  \href{http://dx.doi.org/10.22331/q-2022-09-29-823}{\emph{Quantum} {\bf 6}
  (2022) 823}, [\href{http://arxiv.org/abs/2105.09100}{{\tt 2105.09100}}].

\bibitem{Agliardi:2022ghn}
G.~Agliardi, M.~Grossi, M.~Pellen and E.~Prati, \emph{{Quantum integration of
  elementary particle processes}},
  \href{http://dx.doi.org/10.1016/j.physletb.2022.137228}{\emph{Phys. Lett. B}
  {\bf 832} (2022) 137228}, [\href{http://arxiv.org/abs/2201.01547}{{\tt
  2201.01547}}].

\bibitem{deLejarza:2023qxk}
J.~J.~M. de~Lejarza, M.~Grossi, L.~Cieri and G.~Rodrigo, \emph{{Quantum Fourier
  Iterative Amplitude Estimation}},  in \emph{{2023 International Conference on
  Quantum Computing and Engineering}}, IEEE, 5, 2023.
\newblock \href{http://arxiv.org/abs/2305.01686}{{\tt 2305.01686}}.
\newblock \href{http://dx.doi.org/10.1109/QCE57702.2023.00071}{DOI}.

\bibitem{deLejarza:2024pgk}
J.~J.~M. de~Lejarza, L.~Cieri, M.~Grossi, S.~Vallecorsa and G.~Rodrigo,
  \emph{{Loop Feynman integration on a quantum computer}},
  \href{http://dx.doi.org/10.1103/PhysRevD.110.074031}{\emph{Phys. Rev. D} {\bf
  110} (2024) 074031}, [\href{http://arxiv.org/abs/2401.03023}{{\tt
  2401.03023}}].

\bibitem{deLejarza:2024scm}
J.~J.~M. de~Lejarza, D.~F. Renter\'\i{}a-Estrada, M.~Grossi and G.~Rodrigo,
  \emph{{Quantum integration of decay rates at second order in perturbation
  theory}},  \href{http://arxiv.org/abs/2409.12236}{{\tt 2409.12236}}.

\bibitem{Ramirez-Uribe:2021ubp}
S.~Ram\'\i{}rez-Uribe, A.~E. Renter\'\i{}a-Olivo, G.~Rodrigo, G.~F.~R. Sborlini
  and L.~Vale~Silva, \emph{{Quantum algorithm for Feynman loop integrals}},
  \href{http://dx.doi.org/10.1007/JHEP05(2022)100}{\emph{JHEP} {\bf 05} (2022)
  100}, [\href{http://arxiv.org/abs/2105.08703}{{\tt 2105.08703}}].

\bibitem{Clemente:2022nll}
G.~Clemente, A.~Crippa, K.~Jansen, S.~Ram\'\i{}rez-Uribe, A.~E.
  Renter\'\i{}a-Olivo, G.~Rodrigo et~al., \emph{{Variational quantum
  eigensolver for causal loop Feynman diagrams and directed acyclic graphs}},
  \href{http://dx.doi.org/10.1103/PhysRevD.108.096035}{\emph{Phys. Rev. D} {\bf
  108} (2023) 096035}, [\href{http://arxiv.org/abs/2210.13240}{{\tt
  2210.13240}}].

\bibitem{Strategy:2019vxc}
R.~K. Ellis et~al., \emph{{Physics Briefing Book}: {Input for the European
  Strategy for Particle Physics Update 2020}},
  \href{http://arxiv.org/abs/1910.11775}{{\tt 1910.11775}}.

\bibitem{Gianotti:2002xx}
F.~Gianotti et~al., \emph{{Physics potential and experimental challenges of the
  LHC luminosity upgrade}},
  \href{http://dx.doi.org/10.1140/epjc/s2004-02061-6}{\emph{Eur. Phys. J. C}
  {\bf 39} (2005) 293--333}, [\href{http://arxiv.org/abs/hep-ph/0204087}{{\tt
  hep-ph/0204087}}].

\bibitem{Abada:2019lih}
{\scshape FCC} collaboration, A.~Abada et~al., \emph{{FCC Physics
  Opportunities}: {Future Circular Collider Conceptual Design Report Volume
  1}}, \href{http://dx.doi.org/10.1140/epjc/s10052-019-6904-3}{\emph{Eur. Phys.
  J. C} {\bf 79} (2019) 474}.

\bibitem{Djouadi:2007ik}
{\scshape ILC} collaboration, G.~Aarons et~al., \emph{{International Linear
  Collider Reference Design Report Volume 2: Physics at the ILC}},
  \href{http://arxiv.org/abs/0709.1893}{{\tt 0709.1893}}.

\bibitem{Roloff:2018dqu}
{\scshape CLIC, CLICdp} collaboration, \emph{{The Compact Linear e$^+$e$^-$
  Collider (CLIC): Physics Potential}},
  \href{http://arxiv.org/abs/1812.07986}{{\tt 1812.07986}}.

\bibitem{CEPCStudyGroup:2018ghi}
{\scshape CEPC Study Group} collaboration, M.~Dong et~al., \emph{{CEPC
  Conceptual Design Report: Volume 2 - Physics \& Detector}},
  \href{http://arxiv.org/abs/1811.10545}{{\tt 1811.10545}}.

\bibitem{Heinrich:2020ybq}
G.~Heinrich, \emph{{Collider Physics at the Precision Frontier}},
  \href{http://dx.doi.org/10.1016/j.physrep.2021.03.006}{\emph{Phys. Rept.}
  {\bf 922} (2021) 1--69}, [\href{http://arxiv.org/abs/2009.00516}{{\tt
  2009.00516}}].

\bibitem{Catani:2008xa}
S.~Catani, T.~Gleisberg, F.~Krauss, G.~Rodrigo and J.-C. Winter, \emph{{From
  loops to trees by-passing Feynman's theorem}},
  \href{http://dx.doi.org/10.1088/1126-6708/2008/09/065}{\emph{JHEP} {\bf 09}
  (2008) 065}, [\href{http://arxiv.org/abs/0804.3170}{{\tt 0804.3170}}].

\bibitem{Bierenbaum:2010cy}
I.~Bierenbaum, S.~Catani, P.~Draggiotis and G.~Rodrigo, \emph{{A Tree-Loop
  Duality Relation at Two Loops and Beyond}},
  \href{http://dx.doi.org/10.1007/JHEP10(2010)073}{\emph{JHEP} {\bf 10} (2010)
  073}, [\href{http://arxiv.org/abs/1007.0194}{{\tt 1007.0194}}].

\bibitem{Bierenbaum:2012th}
I.~Bierenbaum, S.~Buchta, P.~Draggiotis, I.~Malamos and G.~Rodrigo,
  \emph{{Tree-Loop Duality Relation beyond simple poles}},
  \href{http://dx.doi.org/10.1007/JHEP03(2013)025}{\emph{JHEP} {\bf 03} (2013)
  025}, [\href{http://arxiv.org/abs/1211.5048}{{\tt 1211.5048}}].

\bibitem{Buchta:2014dfa}
S.~Buchta, G.~Chachamis, P.~Draggiotis, I.~Malamos and G.~Rodrigo, \emph{{On
  the singular behaviour of scattering amplitudes in quantum field theory}},
  \href{http://dx.doi.org/10.1007/JHEP11(2014)014}{\emph{JHEP} {\bf 11} (2014)
  014}, [\href{http://arxiv.org/abs/1405.7850}{{\tt 1405.7850}}].

\bibitem{Buchta:2015wna}
S.~Buchta, G.~Chachamis, P.~Draggiotis and G.~Rodrigo, \emph{{Numerical
  implementation of the loop\textendash{}tree duality method}},
  \href{http://dx.doi.org/10.1140/epjc/s10052-017-4833-6}{\emph{Eur. Phys. J.
  C} {\bf 77} (2017) 274}, [\href{http://arxiv.org/abs/1510.00187}{{\tt
  1510.00187}}].

\bibitem{Hernandez-Pinto:2015ysa}
R.~J. Hernandez-Pinto, G.~F.~R. Sborlini and G.~Rodrigo, \emph{{Towards gauge
  theories in four dimensions}},
  \href{http://dx.doi.org/10.1007/JHEP02(2016)044}{\emph{JHEP} {\bf 02} (2016)
  044}, [\href{http://arxiv.org/abs/1506.04617}{{\tt 1506.04617}}].

\bibitem{Sborlini:2016gbr}
G.~F.~R. Sborlini, F.~Driencourt-Mangin, R.~Hernandez-Pinto and G.~Rodrigo,
  \emph{{Four-dimensional unsubtraction from the loop-tree duality}},
  \href{http://dx.doi.org/10.1007/JHEP08(2016)160}{\emph{JHEP} {\bf 08} (2016)
  160}, [\href{http://arxiv.org/abs/1604.06699}{{\tt 1604.06699}}].

\bibitem{Sborlini:2016hat}
G.~F.~R. Sborlini, F.~Driencourt-Mangin and G.~Rodrigo, \emph{{Four-dimensional
  unsubtraction with massive particles}},
  \href{http://dx.doi.org/10.1007/JHEP10(2016)162}{\emph{JHEP} {\bf 10} (2016)
  162}, [\href{http://arxiv.org/abs/1608.01584}{{\tt 1608.01584}}].

\bibitem{Tomboulis:2017rvd}
E.~T. Tomboulis, \emph{{Causality and Unitarity via the Tree-Loop Duality
  Relation}}, \href{http://dx.doi.org/10.1007/JHEP05(2017)148}{\emph{JHEP} {\bf
  05} (2017) 148}, [\href{http://arxiv.org/abs/1701.07052}{{\tt 1701.07052}}].

\bibitem{Driencourt-Mangin:2017gop}
F.~Driencourt-Mangin, G.~Rodrigo and G.~F.~R. Sborlini, \emph{{Universal dual
  amplitudes and asymptotic expansions for $gg\rightarrow H$ and $H\rightarrow
  \gamma \gamma $ in four dimensions}},
  \href{http://dx.doi.org/10.1140/epjc/s10052-018-5692-5}{\emph{Eur. Phys. J.
  C} {\bf 78} (2018) 231}, [\href{http://arxiv.org/abs/1702.07581}{{\tt
  1702.07581}}].

\bibitem{Jurado:2017xut}
J.~L. Jurado, G.~Rodrigo and W.~J. Torres~Bobadilla, \emph{{From Jacobi
  off-shell currents to integral relations}},
  \href{http://dx.doi.org/10.1007/JHEP12(2017)122}{\emph{JHEP} {\bf 12} (2017)
  122}, [\href{http://arxiv.org/abs/1710.11010}{{\tt 1710.11010}}].

\bibitem{Driencourt-Mangin:2019aix}
F.~Driencourt-Mangin, G.~Rodrigo, G.~F.~R. Sborlini and W.~J. Torres~Bobadilla,
  \emph{{Universal four-dimensional representation of $H \to \gamma \gamma$ at
  two loops through the Loop-Tree Duality}},
  \href{http://dx.doi.org/10.1007/JHEP02(2019)143}{\emph{JHEP} {\bf 02} (2019)
  143}, [\href{http://arxiv.org/abs/1901.09853}{{\tt 1901.09853}}].

\bibitem{Runkel:2019yrs}
R.~Runkel, Z.~Sz\H{o}r, J.~P. Vesga and S.~Weinzierl, \emph{{Causality and
  loop-tree duality at higher loops}},
  \href{http://dx.doi.org/10.1103/PhysRevLett.122.111603}{\emph{Phys. Rev.
  Lett.} {\bf 122} (2019) 111603}, [\href{http://arxiv.org/abs/1902.02135}{{\tt
  1902.02135}}].

\bibitem{Aguilera-Verdugo:2019kbz}
J.~J. Aguilera-Verdugo, F.~Driencourt-Mangin, J.~Plenter,
  S.~Ram\'\i{}rez-Uribe, G.~Rodrigo, G.~F.~R. Sborlini et~al.,
  \emph{{Causality, unitarity thresholds, anomalous thresholds and infrared
  singularities from the loop-tree duality at higher orders}},
  \href{http://dx.doi.org/10.1007/JHEP12(2019)163}{\emph{JHEP} {\bf 12} (2019)
  163}, [\href{http://arxiv.org/abs/1904.08389}{{\tt 1904.08389}}].

\bibitem{Runkel:2019zbm}
R.~Runkel, Z.~Sz\H{o}r, J.~P. Vesga and S.~Weinzierl, \emph{{Integrands of loop
  amplitudes within loop-tree duality}},
  \href{http://dx.doi.org/10.1103/PhysRevD.101.116014}{\emph{Phys. Rev. D} {\bf
  101} (2020) 116014}, [\href{http://arxiv.org/abs/1906.02218}{{\tt
  1906.02218}}].

\bibitem{Capatti:2019ypt}
Z.~Capatti, V.~Hirschi, D.~Kermanschah and B.~Ruijl, \emph{{Loop-Tree Duality
  for Multiloop Numerical Integration}},
  \href{http://dx.doi.org/10.1103/PhysRevLett.123.151602}{\emph{Phys. Rev.
  Lett.} {\bf 123} (2019) 151602}, [\href{http://arxiv.org/abs/1906.06138}{{\tt
  1906.06138}}].

\bibitem{Driencourt-Mangin:2019yhu}
F.~Driencourt-Mangin, G.~Rodrigo, G.~F.~R. Sborlini and W.~J. Torres~Bobadilla,
  \emph{{Interplay between the loop-tree duality and helicity amplitudes}},
  \href{http://dx.doi.org/10.1103/PhysRevD.105.016012}{\emph{Phys. Rev. D} {\bf
  105} (2022) 016012}, [\href{http://arxiv.org/abs/1911.11125}{{\tt
  1911.11125}}].

\bibitem{Capatti:2019edf}
Z.~Capatti, V.~Hirschi, D.~Kermanschah, A.~Pelloni and B.~Ruijl,
  \emph{{Numerical Loop-Tree Duality: contour deformation and subtraction}},
  \href{http://dx.doi.org/10.1007/JHEP04(2020)096}{\emph{JHEP} {\bf 04} (2020)
  096}, [\href{http://arxiv.org/abs/1912.09291}{{\tt 1912.09291}}].

\bibitem{Plenter:2019jyj}
J.~Plenter, \emph{{Asymptotic Expansions Through the Loop-Tree Duality}},
  \href{http://dx.doi.org/10.5506/APhysPolB.50.1983}{\emph{Acta Phys. Polon. B}
  {\bf 50} (2019) 1983--1992}.

\bibitem{Plenter:2020lop}
J.~Plenter and G.~Rodrigo, \emph{{Asymptotic expansions through the loop-tree
  duality}},
  \href{http://dx.doi.org/10.1140/epjc/s10052-021-09094-9}{\emph{Eur. Phys. J.
  C} {\bf 81} (2021) 320}, [\href{http://arxiv.org/abs/2005.02119}{{\tt
  2005.02119}}].

\bibitem{Prisco:2020kyb}
R.~M. Prisco and F.~Tramontano, \emph{{Dual subtractions}},
  \href{http://dx.doi.org/10.1007/JHEP06(2021)089}{\emph{JHEP} {\bf 06} (2021)
  089}, [\href{http://arxiv.org/abs/2012.05012}{{\tt 2012.05012}}].

\bibitem{Verdugo:2020kzh}
J.~J. Aguilera-Verdugo, F.~Driencourt-Mangin, R.~J. Hern\'andez-Pinto,
  J.~Plenter, S.~Ramirez-Uribe, A.~E. Renteria~Olivo et~al., \emph{{Open Loop
  Amplitudes and Causality to All Orders and Powers from the Loop-Tree
  Duality}},
  \href{http://dx.doi.org/10.1103/PhysRevLett.124.211602}{\emph{Phys. Rev.
  Lett.} {\bf 124} (2020) 211602}, [\href{http://arxiv.org/abs/2001.03564}{{\tt
  2001.03564}}].

\bibitem{snowmass2020}
J.~J. Aguilera-Verdugo, R.~J. Hern\'andez-Pinto, S.~Ram\'\i{}rez-Uribe, A.~E.
  Renter\'\i{}a-Olivo, G.~Rodrigo, G.~F.~R. Sborlini et~al., \emph{{Manifestly
  Causal Scattering Amplitudes}}, {\emph{Snowmass LoI} (August 2020) }.

\bibitem{Aguilera-Verdugo:2020kzc}
J.~J. Aguilera-Verdugo, R.~J. Hernandez-Pinto, G.~Rodrigo, G.~F.~R. Sborlini
  and W.~J. Torres~Bobadilla, \emph{{Causal representation of multi-loop
  Feynman integrands within the loop-tree duality}},
  \href{http://dx.doi.org/10.1007/JHEP01(2021)069}{\emph{JHEP} {\bf 01} (2021)
  069}, [\href{http://arxiv.org/abs/2006.11217}{{\tt 2006.11217}}].

\bibitem{Aguilera-Verdugo:2020nrp}
J.~Jes\'us Aguilera-Verdugo, R.~J. Hern\'andez-Pinto, G.~Rodrigo, G.~F.~R.
  Sborlini and W.~J. Torres~Bobadilla, \emph{{Mathematical properties of nested
  residues and their application to multi-loop scattering amplitudes}},
  \href{http://dx.doi.org/10.1007/JHEP02(2021)112}{\emph{JHEP} {\bf 02} (2021)
  112}, [\href{http://arxiv.org/abs/2010.12971}{{\tt 2010.12971}}].

\bibitem{Ramirez-Uribe:2020hes}
S.~Ram\'\i{}rez-Uribe, R.~J. Hern\'andez-Pinto, G.~Rodrigo, G.~F.~R. Sborlini
  and W.~J. Torres~Bobadilla, \emph{{Universal opening of four-loop scattering
  amplitudes to trees}},
  \href{http://dx.doi.org/10.1007/JHEP04(2021)129}{\emph{JHEP} {\bf 04} (2021)
  129}, [\href{http://arxiv.org/abs/2006.13818}{{\tt 2006.13818}}].

\bibitem{Sborlini:2021owe}
G.~F.~R. Sborlini, \emph{{Geometrical approach to causality in multiloop
  amplitudes}},
  \href{http://dx.doi.org/10.1103/PhysRevD.104.036014}{\emph{Phys. Rev. D} {\bf
  104} (2021) 036014}, [\href{http://arxiv.org/abs/2102.05062}{{\tt
  2102.05062}}].

\bibitem{TorresBobadilla:2021ivx}
W.~J. Torres~Bobadilla, \emph{{Loop-tree duality from vertices and edges}},
  \href{http://dx.doi.org/10.1007/JHEP04(2021)183}{\emph{JHEP} {\bf 04} (2021)
  183}, [\href{http://arxiv.org/abs/2102.05048}{{\tt 2102.05048}}].

\bibitem{TorresBobadilla:2021dkq}
W.~J.~T. Bobadilla, \emph{{Lotty \textendash{} The loop-tree duality
  automation}},
  \href{http://dx.doi.org/10.1140/epjc/s10052-021-09235-0}{\emph{Eur. Phys. J.
  C} {\bf 81} (2021) 514}, [\href{http://arxiv.org/abs/2103.09237}{{\tt
  2103.09237}}].

\bibitem{Aguilera-Verdugo:2021nrn}
J.~de~Jes\'us Aguilera-Verdugo et~al., \emph{{A Stroll through the Loop-Tree
  Duality}}, \href{http://dx.doi.org/10.3390/sym13061029}{\emph{Symmetry} {\bf
  13} (2021) 1029}, [\href{http://arxiv.org/abs/2104.14621}{{\tt 2104.14621}}].

\bibitem{Rios-Sanchez:2024xtv}
J.~Rios-Sanchez and G.~Sborlini, \emph{{Toward multiloop local renormalization
  within causal loop-tree duality}},
  \href{http://dx.doi.org/10.1103/PhysRevD.109.125004}{\emph{Phys. Rev. D} {\bf
  109} (2024) 125004}, [\href{http://arxiv.org/abs/2402.13995}{{\tt
  2402.13995}}].

\bibitem{Ramirez-Uribe:2024rjg}
S.~Ram\'\i{}rez-Uribe, P.~K. Dhani, G.~F.~R. Sborlini and G.~Rodrigo,
  \emph{{Rewording Theoretical Predictions at Colliders with Vacuum
  Amplitudes}},
  \href{http://dx.doi.org/10.1103/PhysRevLett.133.211901}{\emph{Phys. Rev.
  Lett.} {\bf 133} (2024) 211901}, [\href{http://arxiv.org/abs/2404.05491}{{\tt
  2404.05491}}].

\bibitem{LTD:2024yrb}
{\scshape LTD} collaboration, S.~Ram\'\i{}rez-Uribe, A.~E. Renter\'\i{}a-Olivo,
  D.~F. Renter\'\i{}a-Estrada, J.~J.~M. de~Lejarza, P.~K. Dhani, L.~Cieri
  et~al., \emph{{Vacuum amplitudes and time-like causal unitary in the
  loop-tree duality}},  \href{http://arxiv.org/abs/2404.05492}{{\tt
  2404.05492}}.

\bibitem{Even20111}
S.~Even and G.~Even, \emph{Graph Algorithms, second edition},
  vol.~9780521517188.
\newblock 2011.
\newblock 10.1017/CBO9781139015165.

\bibitem{Grover:1997fa}
L.~K. Grover, \emph{{Quantum mechanics helps in searching for a needle in a
  haystack}}, \href{http://dx.doi.org/10.1103/PhysRevLett.79.325}{\emph{Phys.
  Rev. Lett.} {\bf 79} (1997) 325--328},
  [\href{http://arxiv.org/abs/quant-ph/9706033}{{\tt quant-ph/9706033}}].

\bibitem{Boyer:1996zf}
M.~Boyer, G.~Brassard, P.~Hoeyer and A.~Tapp, \emph{{Tight bounds on quantum
  searching}},
  \href{http://dx.doi.org/10.1002/(SICI)1521-3978(199806)46:4/5<493::AID-PROP493>3.0.CO;2-P}{\emph{Fortsch.
  Phys.} {\bf 46} (1998) 493--506},
  [\href{http://arxiv.org/abs/quant-ph/9605034}{{\tt quant-ph/9605034}}].

\bibitem{EPpatent}
S.~Ram\'\i{}rez-Uribe, A.~E. Renter\'\i{}a-Olivo, G.~Rodrigo and G.~F.~R.
  Sborlini, \emph{{Method for generating in a random order directed acyclic
  graph (DAG) configurations of a given graph in a quantum computing device }},
  {\emph{European patent application EP24382213} (28 February 2024) }.

\bibitem{Bollini:1972ui}
C.~G. Bollini and J.~J. Giambiagi, \emph{{Dimensional Renormalization: The
  Number of Dimensions as a Regularizing Parameter}},
  \href{http://dx.doi.org/10.1007/BF02895558}{\emph{Nuovo Cim. B} {\bf 12}
  (1972) 20--26}.

\bibitem{tHooft:1972tcz}
G.~'t~Hooft and M.~J.~G. Veltman, \emph{{Regularization and Renormalization of
  Gauge Fields}},
  \href{http://dx.doi.org/10.1016/0550-3213(72)90279-9}{\emph{Nucl. Phys. B}
  {\bf 44} (1972) 189--213}.

\bibitem{Nielsen:2012yss}
M.~A. Nielsen and I.~L. Chuang, \emph{{Quantum Computation and Quantum
  Information}}.
\newblock Cambridge University Press, 6, 2012.
\newblock 10.1017/cbo9780511976667.

\bibitem{Depth:5769}
\emph{https://quantumcomputing.stackexchange.com/questions/5769/how-to-calculate-circuit-depth-properly},
  2020.

\end{thebibliography}\endgroup

\end{document}